\documentclass[12pt]{article}
\usepackage{axodraw}
\usepackage{a4wide,epsfig,psfrag,amsmath,amssymb,cite,scalefnt,color}

\newcommand{\gsim}{\;\rlap{\lower 3.5 pt \hbox{$\mathchar \sim$}} \raise 1pt
 \hbox {$>$}\;}
\newcommand{\lsim}{\;\rlap{\lower 3.5 pt \hbox{$\mathchar \sim$}} \raise 1pt
 \hbox {$<$}\;}

\begin{document}


\title{\vskip-3cm{\baselineskip14pt
    \begin{flushleft}
      \normalsize SFB/CPP-09-86\\
      \normalsize TTP09-33
  \end{flushleft}}
  \vskip1.5cm
  Gluinonia: Energy Levels, Production and Decay
}
\author{\small Matthias R. Kauth, Johann H. K\"uhn, Peter Marquard and Matthias Steinhauser\\[1em]
  {\small\it Institut f{\"u}r Theoretische Teilchenphysik}\\
  {\small\it Karlsruhe Institute of Technology (KIT)}\\
  {\small\it 76128 Karlsruhe, Germany}}

\date{}

\maketitle

\thispagestyle{empty}

\begin{abstract}

Predictions for energy levels, production and decay rate of gluinonia,
non-relativistic boundstates of gluinos, are presented. The potential
between color-octet constituents is derived in next-to-next-to leading
order and one-loop QCD corrections are derived for the production cross
section and the decay rate into gluon jets. In addition we
evaluate the decay rate into top quarks and into two photons. The
signal-to-background ratio is estimated for the dominant decay mode and found  
to be around $0.5\%$. For relatively light gluinos the bound states thus
might be detectable.

\medskip



\end{abstract}

\thispagestyle{empty}


\newpage

\section{\label{sec:introduction}Introduction}

With the turning-on of the LHC in the near future, one of the most
important tasks of both general-purpose detectors ATLAS and CMS will be
the search for physics beyond the Standard Model, with one of the most
promising candidates being Supersymmetry.  In the course of this
program, which will obviously be adjusted and focused with increasing
luminosity and with the first indications of ``New Physics'', the
determination of the properties of newly detected particles will be a
major task. Indeed, only after the quantum numbers, the interaction and
the mass of a new particle have been measured with high precision, this
requirement can be considered fulfilled.

In the present paper we want to concentrate on the gluino, the
supersymmetric partner of the gluon, and discuss the possibility to
determine its quantum numbers, i.e. its spin, color, Majorana nature and
its mass, through the investigation of non-relativistic boundstates of
two gluinos, generally denoted ``gluinonia''.  The classification of
gluinonia according to their quantum numbers and their qualitative
properties, i.e. spectra and decay modes, as well as estimates of their
production cross section at hadron colliders have been studied already
more than two decades ago and presented in
Refs.~\cite{Keung:1983wz,Kuhn:1983sc,Goldman:1984mj} concentrating at
that time on gluino masses which are by now experimentally excluded.
More recently gluinonia have been discussed in
Refs.~\cite{Kartvelishvili:1989pp,Chikovani:1996bk}. Hadronic
transitions between different gluinonium levels and some rare
annihilation decays were evaluated in \cite{Kartvelishvili:1989pp}.  The
production of vector and pseudoscalar states at the Tevatron and the LHC
was investigated in \cite{Chikovani:1996bk} where it was argued that,
given sufficiently good yet realistic jet mass resolution, pseudoscalar
states could be detected for bound state masses up to 3 TeV. A
phenomenological study of the question, to which extent a signal for
color-octet states could indeed be detected experimentally, has been
performed in \cite{Kartvelishvili:1988tm}. In all these cases, however,
the production cross section has been calculated in lowest order
approximation only and a qualitative phenomenological potential for the
gluino-gluino interaction has been employed. The topic has gained
renewed interest in connection with Split SUSY \cite{ArkaniHamed:2004fb,
  Giudice:2004tc}, (see, e.g.,
Refs.~\cite{Kilian:2004uj,Cheung:2004ad}).

The detection of these bound states would provide important information
about gluino properties, difficult to obtain through other means. Since
gluinonia only exist if gluinos do not decay too rapidly, say with decay
rates less than a few GeV, their observation would immediately provide
at least an upper limit on $\Gamma_{\tilde g}$.  The energy spectrum of
the gluinonium system, e.g. the mass difference between ground state and
first radial excitation (or open gluino production) would be sensitive
to the potential and thus to the color-octet nature of the constituents.
The same is true for the production cross section which is proportional
to the square of the wave function at the origin.  Gluinos are expected
to decay through cascades into several jets plus the lightest
supersymmetric particle (LSP) which escapes detection.  For this reason
the precise determination of its mass in direct gluino decays is limited
by the missing mass or energy resolution.

In contrast, once the mass of the boundstate and the dynamics of the
interaction are known with sufficient precision, the mass of the
constituents is fixed with the corresponding accuracy and in any case
subject to a different systematic error.

The energy levels of gluinonia were estimated already in the early
papers employing phenomenological color-octet potentials, obtained from
phenomenological quarkonium potentials by multiplication with the ratio
of the corresponding Casimir operators $C_{A}/C_{F}= 9/4$ and the gluino
mass was interpreted as an effective mass closely related to the pole
mass. During recent years more refined studies have been performed to
describe non-relativistic boundstates of color-triplet constituents. This
includes the evaluation of the static two-loop potential
\cite{Peter:1996ig,Peter:1997me,Schroder:1998vy}, first steps toward the
evaluation of the three-loop potential \cite{Smirnov:2008pn} and the
calculation of mass-suppressed terms proportional $\alpha_{s}^2/m$
\cite{Gupta:1981pd,Gupta:1982,Titard:1993nn,Kniehl:2001ju,Pineda:1997hz} and
$\alpha_{s}/m^2$
\cite{Pineda:1997bj,Manohar:2000hj,Titard:1993nn,Pineda:1997hz}. Mass
definitions better suited to control the long distance part of the
potential have been introduced \cite{Beneke:1998rk,Hoang:1998nz}, and
the evaluation of the spectra through analytic methods has been
possible, using a perturbative approach to include higher order
contributions from the potential
\cite{Kuhn:1998uy,Penin:2002zv,Kniehl:2002br}. Last but not least,
finite-width effects were incorporated for top-quark production at
electron-positron \cite{Fadin:1987wz,Kuhn:1987ty,Strassler:1990nw,
  Sumino:1992ai,Jezabek:1992np,Hoang:2000yr,Hoang:2004tg} and hadron colliders
\cite{Fadin:1990wx,Hagiwara:2008df,Kiyo:2008bv}. These new results can
be applied also in studies of gluino bound states.

In order to arrive at reliable predictions for the signals of gluinonium
production and decay, the inclusion of a realistic QCD potential and
next-to-leading order (NLO) corrections to boundstate production at the
LHC is mandatory. Considering and building on the significant progress
in the perturbative treatment of non-relativistic boundstates, in
particular quarkonium and positronium, predictions of similar quality
can be made for boundstates with color-octet constituents. This is the
main content of the present paper.

We start with a brief overview and a discussion of the quantum numbers
and qualitative properties of boundstates with Majorana constituents and
evaluate the QCD potential for a color-octet boundstate (Section
\ref{sec:properties}). Subsequently, in Section \ref{sec:spectroscopy},
we calculate the spectrum and the wave function at the origin for the
lowest-lying levels.

These quantities are the necessary ingredients for a realistic
prediction of gluinonium production and decay rates presented in Section
\ref{sec:production}. Assuming that squarks are relatively heavy, the
dominant decay channel of the pseudoscalar singlet boundstate proceeds
through two gluons, and it will be a difficult task to distinguish this
signal from the irreducible background from the two-gluon jet
continuum.

In Section \ref{sec:photons} we consider as an alternative the decays
into top quarks and two photons, which can proceed through virtual quark ($q$)
and 
squark ($\tilde q$) intermediate states. One might, in principle, hope
that the extremely clean $\gamma\gamma$ signal sticks out of the irreducible
background similarly to the Higgs decay into two photons
which has been identified as a promising signal significantly superior
to the $b\bar b$ or two-gluon mode.
Similar considerations apply to the $t\bar{t}$ mode.

In order that these modes can compete with the two-gluon channel, squarks
should not be dramatically heavier than gluinos. On the other hand, to
avoid dominance of the single-gluino decay, the decay mode $\tilde g \to
\tilde q \bar q$ should be kinematically forbidden, corresponding to
$m_{\tilde g}\le m_{\tilde q}$. We therefore evaluate the corresponding
decay rates and study if, given suitable choices for the respective
masses, the two-photon mode might lead to a possible signal. In
Section~\ref{sec:background} a rough estimate of the dominant signal versus 
background is presented. Section~\ref{sec:conclusions} contains our
conclusions.


\section{\label{sec:properties}Properties of gluinonia}
Let us start with a brief recapitulation of the quantum numbers of gluino
boundstates and the corresponding color, spin and orbital momentum
configurations \cite{Keung:1983wz,Kuhn:1983sc,Goldman:1984mj}. These
differ from those of quarkonia due to the restrictions arising from the
Majorana nature of gluinos, and due to their different color assignment.

Two color-octet states can be combined into irreducible representations
as follows
\begin{eqnarray}
8\otimes 8&=&1_{s}\oplus 8_{s}\oplus 8_{a}\oplus
10_{a}\oplus\overline{10}_{a}\oplus 27_{s}
\,,
\label{8times8}
\end{eqnarray}
where the index indicates the (anti-)symmetry with respect to their
color index. The interaction can be either attractive, repulsive or
absent (in lowest order). In lowest order the coefficient of the QCD
potential is given by the expectation value of the product of the color
generators $F^{a}_{ij}F^{a}_{kl}$,
taken between two-particle states in the respective representation.

This product, in turn, can be expressed by the eigenvalues of the
quadratic Casimir operator of the constituents, $C_{A}=3$, and the
boundstate in representation $R$, $C_{R}=\left(F^{R}\right)^2$:
\begin{eqnarray}
  F^{a,1}\cdot F^{a,2}&=&
  \frac{1}{2}\left[(F^{R})^2-(F^{a,1})^2-(F^{a,2})^2\right]=\frac{1}{2}\left(C_{{R}}-2C_{A}\right)
  \,.
\label{FdotF}
\end{eqnarray}
The results are listed in Tab.~\ref{table1}.
In the following we shall limit the discussion to the cases with 
negative coefficients, corresponding to attraction.

\begin{table}[t]
\begin{center}
\begin{tabular}{c|ccc}
R & $(F^{R})^2$ & $F^{a,1}\cdot F^{a,2}$ & \mbox{interaction} \\ \hline
$1_{s}$ & $0$ & $-3$ & \mbox{attractive}\\
$8_{s},8_{a}$ & $3$ & $-\frac{3}{2}$ & \mbox{attractive}\\
$10_{a},\overline{10}_{a}$ & $6$ & $0$ & \mbox{neutral}\\
$27_{s}$ & $8$ & $1$ & \mbox{repulsive}
\end{tabular}
\caption{Color interaction of two $\mbox{SU}(3)$ octets.}
\label{table1}
\end{center}
\end{table}

Fermi statistics and the Majorana nature of the gluinos lead to
additional restrictions. For the symmetric color configurations $1_s$
and $8_s$ antisymmetric spin-angular momentum wave functions,
$(-1)^{L+S}=1$, are required, for the antisymmetric color configuration
$8_{a}$ symmetric ones, $(-1)^{L+S}=-1$. The intrinsic parity of a
Majorana particle can be chosen to be imaginary, its  parity under
charge reflection real, leading to negative intrinsic parity and
positive charge parity of the boundstate. For a few lowest orbital
angular momenta the complete set of boundstate quantum numbers is listed
in Tab.~\ref{table2}.

\begin{table}[t]
\begin{center}
\begin{tabular}{c|ccccccc}
$^{2S+1}L_{J}$ & $^{1}S_{0}$ & $^{3}S_{1}$ & $^{1}P_{1}$ & $^{3}P_{0}$
& $^{3}P_{1}$ & $^{3}P_{2}$ & $^{1}D_{2}$ \\ \hline
L & $0$ & $0$ & $1$ & $1$ & $1$ & $1$ & $2$ \\
S & $0$ & $1$ & $0$ & $1$ & $1$ & $1$ & $0$ \\
$(\tilde{g}\tilde{g})_{s}$ & $0^{-+}$ & $-$ & $-$ & $0^{++}$ & $1^{++}$ & $2^{++}$ & $2^{-+}$ \\
$(\tilde{g}\tilde{g})_{a}$ & $-$ & $1^{-+}$ & $1^{++}$ & $-$ & $-$ & $-$ & $-$
\end{tabular}
\caption{Lowest-lying states $J^{PC}$ of the gluinonium spectrum.}
\label{table2}
\end{center}
\end{table}

The production rate of non-relativistic boundstates in hard collisions 
is proportional to the squared wave function at the origin (for $S$ waves) 
or its derivative (for $P$ waves), the latter being significantly
suppressed and of relative order $v^2$. For this reason the following
discussion will be limited to $S$ waves only. We, furthermore,
anticipate that the experimental mass resolution and, eventually, the
large natural width of the boundstate  will lead to a sizable smearing
of the ``narrow'' resonances. The signal to background ratio will be
small and their detection difficult. For this reason we
will consider color-singlet states only, where the level spacings and
production rates are enhanced approximately by the square of the color
coefficient listed in Tab.~\ref{table1}.

Let us now discuss the QCD potential. Similarly to the case of heavy-quark
boundstates it can be decomposed into the following two terms
\begin{eqnarray}
V_{\rm \tilde{g}\tilde{g}}(\vec{r}\,) & = &  V_{\rm C}(r) + V_{\rm nC}(\vec{r}\,)\,,
\label{pot}
\end{eqnarray}
which are ordered according to their inverse powers in the constituent
mass  $m$ and are given in coordinate space in close similarity to the
quarkonium potential (see e.g. \cite{Pineda:1997hz}). For $S$ waves it
reads
\begin{eqnarray}
V_{\rm C}(r)&=&-\frac{C_{A}\alpha_{s}(\mu^2)}{r}\biggl\{1+\frac{\alpha_{s}(\mu^2)}{4\pi}\biggl[a_{1}+2\gamma_{E}\beta_{0}+2\beta_{0}\ln(\mu r)\biggr]\biggr.
\nonumber\\
&&\biggl.+\frac{\alpha_{s}(\mu^2)^2}{(4\pi)^2}\biggl[a_{2}+2\gamma_{E}\left(2 a_{1}\beta_{0}+\beta_{1}\right)+\beta_{0}^2\left(\frac{\pi^2}{3}+4\gamma_{E}^2\right)\biggr.\biggr.
\nonumber\\
&&\biggl.\biggl.+\left(4a_{1}\beta_{0}+2\beta_{1}+8\gamma_{E}\beta_{0}^2\right)\ln(\mu
r)+4\beta_{0}^2\ln(\mu r)^2\biggr]\biggr\}
\,,\nonumber\\
V_{\rm nC}(\vec{r}\,)&=&U_{1}(r)+U_{2}(\vec{r}\,)+U_{3}(r)
\,,\nonumber\\
U_{1}(r)&=&-\frac{\Delta^2}{4m_{\tilde{g}}^3}+\frac{C_{A}\alpha_{s}(\mu^2)}{m_{\tilde{g}}^2 r}\Delta
\nonumber\,,\\
U_{2}(\vec{r}\,)&=&\frac{4\pi
  C_{A}\alpha_{s}(\mu^2)}{3m_{\tilde{g}}^2}\vec{S}^2\delta\left(\vec{r}\,\right)
\,,\nonumber\\
U_{3}(r)&=&-\frac{C_{A}^2\alpha_{s}^2(\mu^2)}{4m_{\tilde{g}} r^2}
\,,\label{VundU}
\end{eqnarray}
where
\begin{eqnarray}
\beta_{0}&=&\frac{11}{3}C_{A}-\frac{4}{3}T_{F}n_{f}
\,,\nonumber\\
\beta_{1}&=&\frac{34}{3}C_{A}^2-\frac{20}{3}C_{A}T_{F}n_{f}-4C_{F}T_{F}n_{f}
\,,\nonumber\\
a_{1}&=&\frac{31}{9}C_{A}-\frac{20}{9}T_{F}n_{f}
\,,\nonumber\\
a_{2}&=&\left[\frac{4343}{162}+4
  \pi^2-\frac{\pi^4}{4}+\frac{22}{3}\zeta(3) \right]
C_{A}^2-\left[\frac{1798}{81}+\frac{56}{3}\zeta(3)\right]C_{A}T_{F}n_{f}
\nonumber\\
&&-\left[\frac{55}{3}-16\zeta(3)\right]C_{F}T_{F}n_{f}+\left(\frac{20}{9}T_{F}n_{f}\right)^2
\,.\label{BetaundA}
\end{eqnarray}
Here $\alpha_s$ stands for the strong coupling constant in the $\overline{\mbox{MS}}$
scheme, $n_{f}$ is the number of active quark flavors, $C_{A}=3$, $C_{F}=4/3$,
$T_{F}=1/2$, $\gamma_E=0.5772\dots$ the Euler constant,
$\zeta(3)=1.2021\dots$ and $m_{\tilde{g}}$ is the constituent pole mass.

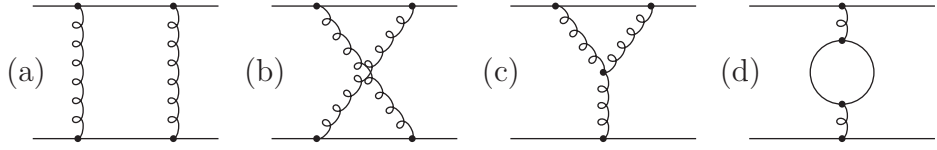
\begin{figure}[t]
\begin{center}
\begin{picture}(350,50)(0,0)
\SetColor{Black}
\Text(0,25)[l]{(a)}
\Line(10,50)(80,50)
\Vertex(27,50){1.3}
\Vertex(63,50){1.3}
\Gluon(27,50)(27,0){2}{6}
\Gluon(63,50)(63,0){2}{6}
\Vertex(27,0){1.3}
\Vertex(63,0){1.3}
\Line(10,0)(80,0)
\Text(90,25)[l]{(b)}
\Line(100,50)(170,50)
\Vertex(117,50){1.3}
\Vertex(153,50){1.3}
\Gluon(117,50)(153,0){2}{8}
\Gluon(153,50)(117,0){2}{8}
\Vertex(117,0){1.3}
\Vertex(153,0){1.3}
\Line(100,0)(170,0)
\Text(180,25)[l]{(c)}
\Line(190,50)(260,50)
\Vertex(207,50){1.3}
\Vertex(243,50){1.3}
\Gluon(207,50)(225,25){2}{4}
\Gluon(243,50)(225,25){2}{4}
\Vertex(225,25){1.3}
\Gluon(225,25)(225,0){2}{3}
\Vertex(225,0){1.3}
\Line(190,0)(260,0)
\Text(270,25)[l]{(d)}
\Line(280,50)(350,50)
\Vertex(315,50){1.3}
\Gluon(315,50)(315,37){2}{1}
\Vertex(315,37){1.3}
\CArc(315,25)(12,0,360)
\Vertex(315,13){1.3}
\Gluon(315,13)(315,0){2}{1}
\Vertex(315,0){1.3}
\Line(280,0)(350,0)
\end{picture}
\caption{Typical NLO contributions to the $q\overline{q}$ potential.
Straight and curely lines represent quarks and gluons, respectively.}
\label{feynNLO}
\end{center}
\end{figure}

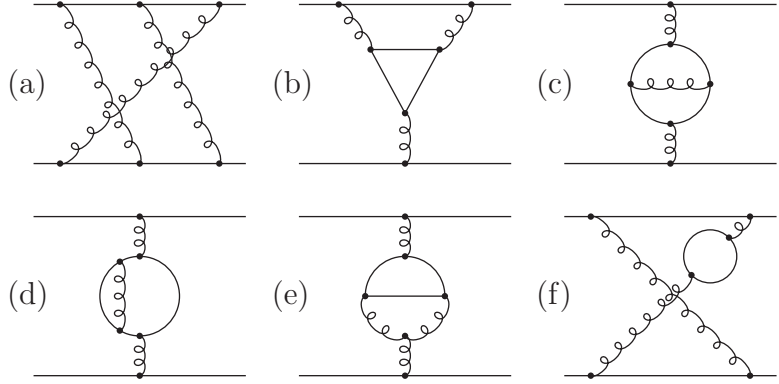
\begin{figure}[t]
\begin{center}
\begin{picture}(280,140)(0,0)
\SetColor{Black}
\Text(-10,110)[l]{(a)}
\Line(0,140)(80,140)
\Vertex(10,140){1.2}
\Vertex(40,140){1.2}
\Vertex(70,140){1.2}
\Gluon(10,140)(40,80){2}{7}
\Gluon(40,140)(70,80){2}{7}
\Gluon(70,140)(10,80){2}{9}
\Vertex(10,80){1.2}
\Vertex(40,80){1.2}
\Vertex(70,80){1.2}
\Line(0,80)(80,80)
\Text(90,110)[l]{(b)}
\Line(100,140)(180,140)
\Vertex(115,140){1.2}
\Vertex(165,140){1.2}
\Gluon(165,140)(153,123){2}{2}
\Gluon(115,140)(127,123){2}{2}
\Vertex(153,123){1.2}
\Vertex(127,123){1.2}
\Vertex(140,99){1.2}
\Line(153,123)(127,123)
\Line(127,123)(140,99)
\Line(140,99)(153,123)
\Gluon(140,99)(140,80){2}{2}
\Vertex(140,80){1.2}
\Line(100,80)(180,80)
\Text(190,110)[l]{(c)}
\Line(200,140)(280,140)
\Vertex(240,140){1.2}
\Gluon(240,140)(240,125){2}{2}
\Vertex(240,125){1.2}
\CArc(240,110)(15,0,360)
\Vertex(225,110){1.2}
\Gluon(255,110)(225,110){2}{3}
\Vertex(255,110){1.2}
\Vertex(240,95){1.2}
\Gluon(240,95)(240,80){2}{2}
\Vertex(240,80){1.2}
\Line(200,80)(280,80)
\Text(-10,30)[l]{(d)}
\Line(0,60)(80,60)
\Vertex(40,60){1.2}
\Gluon(40,60)(40,45){2}{2}
\Vertex(40,45){1.2}
\CArc(40,30)(15,0,360)
\Vertex(32.5,42.99){1.2}
\Gluon(32.5,42.99)(32.5,17.01){2}{3}
\Vertex(32.5,17.01){1.2}
\Vertex(40,15){1.2}
\Gluon(40,15)(40,0){2}{2}
\Vertex(40,0){1.2}
\Line(0,0)(80,0)
\Text(90,30)[l]{(e)}
\Line(100,60)(180,60)
\Vertex(140,60){1.2}
\Gluon(140,60)(140,45){2}{2}
\Vertex(140,45){1.2}
\CArc(140,30)(15,0,180)
\Line(155,30)(125,30)
\Vertex(155,30){1.2}
\Vertex(125,30){1.2}
\GlueArc(140,30)(15,180,270){2}{2}
\GlueArc(140,30)(15,270,360){2}{2}
\Vertex(140,15){1.2}
\Gluon(140,15)(140,0){2}{2}
\Vertex(140,0){1.2}
\Line(100,0)(180,0)
\Text(190,30)[l]{(f)}
\Line(200,60)(280,60)
\Vertex(210,60){1.2}
\Gluon(262.07,52.07)(270,60){-2}{1}
\Gluon(210,0)(247.93,37.93){-2}{6}
\Vertex(270,60){1.2}
\CArc(255,45)(10,0,360)
\Vertex(262.07,52.07){1.2}
\Vertex(247.93,37.93){1.2}
\Vertex(210,0){1.2}
\Gluon(270,0)(210,60){-2}{9}
\Vertex(270,0){1.2}
\Line(200,0)(280,0)
\end{picture}
\caption{Typical NNLO contributions to the $q\overline{q}$
  potential. The same notation as in Fig.~\ref{feynNLO} has been
  adopted.}
\label{feynNNLO}
\end{center}
\end{figure}

For the  Coulomb part $V_{\rm C}$ the transformation from color-triplet, where
the result can be found in the literature, to constituents in an
arbitrary representation $R$, bound to a singlet state, can be
understood as follows: In LO the overall normalization is changed by
substituting $C_{F} \to C_{R}$. This is the only modification also in
NLO, thus leaving the constant $a_{1}$ and the terms proportional to
$\beta_0$ unchanged. For the case under discussion ($R=A$) this is
evident from Fig.~\ref{feynNLO}: Diagram (a) contributes proportional
$C_{R}^2$, diagram (b) proportional $C_{R}^2-\frac{1}{2}C_{R}C_{A}$. Its
$C_{R}^2$ part can be combined with (a) and in total the ladder plus
crossed ladder can be collected in the $C_{R}/r$ part of the
potential. The $C_{R}C_{A}$ part of diagram (b) contributes to the
$C_{A}$ term in $\beta_{0}$ and in the constant $a_{1}$. Diagram (d)
contributes proportional $C_{R}T_{F}n_{f}$ and is responsible for the
$T_{F}n_{f}$ terms in $\beta_{0}$ and $a_{1}$.
The corresponding considerations are also applicable to the 
two-loop case, see e.g. Fig.~\ref{feynNNLO}.
For diagrams with virtual gluons only (e.g. Fig.~\ref{feynNNLO} (a)),
the substitution $C_F \to C_R$ is valid throughout. For diagrams with
exactly one dressed gluon propagator exchange (e.g. Fig.~\ref{feynNNLO}
(c), (d), (e)), the same substitution is valid, as far as the coupling
of the gluon to the constituent is concerned and only the external
$C_{F}$ is replaced by $C_{R}$. Diagrams with three-gluon coupling
through a triangular fermion loop (Fig.~\ref{feynNNLO} (b)) are also
proportional to $C_R C_A T_F n_f$ and it is again the first factor,
which is chosen to be $C_F$ in the case of quark- and $C_A$ in the case
of gluino-constituents. Diagrams similar to those of Fig.~\ref{feynNLO}
(a)-(c), but with one dressed gluon (e.g. Fig.~\ref{feynNNLO} (f)) can
be handled like the NLO diagrams. In total, up to NNLO, the Coulomb part
of the potential for the binding of constituents in an arbitrary
representation $R$ is obtained through the substitution $C_F \to C_R$ in
the overall factor.\footnote{We do not expect that this statement holds
  true to arbitrary high orders. In contrast to heavy quarks gluinos
  together with gluons can form color-singlet states and even in
  quenched QCD the static potential between two gluinos is
  not expected to be confining.}

The term $U_{3}$ of Eq.~(\ref{VundU}) of order $\alpha_s^2 / m$ originates
from a one-loop calculation, and the factor $C_F(C_F/2 - C_A)$ must be
replaced by $C_R(C_R/2 - C_A)$. The terms of order $\alpha_s/m^2$
collected in $U_{1}$ and $U_{2}$ can be translated simply by replacing
the global factor $C_{F}$ by $C_{R}$.
For our application the substitution $C_{R}\to C_{A}$ is implicit.


\section{\label{sec:spectroscopy}Spectroscopy}
Using the potential as specified in Eqs.~(\ref{pot}) and (\ref{VundU}),
binding energies, level spacings and bound-state wave functions are
easily obtained by solving Schr\"odinger's equation numerically for
$V_{C}$ and adding the singular terms collected in $U_{1},U_{2}$
and $U_{3}$ in perturbation theory or, alternatively, by evaluating
energy level and wave functions in a perturbative series in
$\alpha_{s}$.

The analytic results both for binding energies $E_{n}$ and wave
functions at the origin $\left|\Psi_{n}\left(0\right)\right|^2$ are
listed in Appendix~\ref{app:energy}. In numerical form the results for
the two lowest levels are given by
\begin{eqnarray}
E_{1}&\hspace{-0.1cm}=&\hspace{-0.2cm}-\frac{m_{\tilde{g}}C_{A}^2\alpha_{s}^2}{4}\left\{1+\alpha_{s}\left(2.44L_{1}+3.20\right)\right.
\nonumber\\
&&\hspace{0cm}\hspace{1.95cm}\left.+\alpha_{s}^2\left[\left(4.47L_{1}^2+9.71L_{1}+12.47\right)_{C}+\left(20.81\right)_{nC}\right]\right\}
\,,\nonumber\\
E_{2}&\hspace{-0.1cm}=&\hspace{-0.2cm}-\frac{m_{\tilde{g}}C_{A}^2\alpha_{s}^2}{16}\left\{
1+\alpha_{s}\left(2.44L_{2}+4.42\right)\right.
\nonumber\\
&&\hspace{0cm}\hspace{1.95cm}\left.+\alpha_{s}^2\left[\left(4.47L_{2}^2+14.19L_{2}+20.54\right)_{C}+\left(11.95\right)_{nC}\right]\right\}
\,,\nonumber\\
\left|\Psi_{1}\left(0\right)\right|^2&\hspace{-0.1cm}=&\hspace{-0.2cm}\frac{m_{\tilde{g}}^3C_{A}^3\alpha_{s}^3}{8\pi}\left\{
1+\alpha_{s}\left(3.66L_{1}-0.43\right)\right.
\nonumber\\
&&\hspace{-0.2cm}\hspace{1.8cm}\left.+\alpha_{s}^2\left[\left(8.93L_{1}^2-5.11L_{1}+5.83\right)_{C}+\left(27L_{1}+57.38\right)_{nC}\right]\right\}
\,,\nonumber\\
\left|\Psi_{2}\left(0\right)\right|^2&\hspace{-0.1cm}=&\hspace{-0.2cm}\frac{m_{\tilde{g}}^3C_{A}^3\alpha_{s}^3}{64\pi}\left\{
1+\alpha_{s}\left(3.66L_{2}-0.18\right)\right.
\nonumber\\
&&\hspace{-0.2cm}\hspace{1.8cm}\left.+\alpha_{s}^2\left[\left(8.93L_{2}^2-3.86L_{2}+10.19\right)_{C}+\left(27L_{2}+29.53\right)_{nC}\right]\right\}
\,,\label{EundPsiNum}
\end{eqnarray}
with $L_{n}=\ln\left(n\mu/(m_{\tilde{g}}C_{A}\alpha_{s})\right)$ and
$m_{\tilde{g}}$ being the gluino pole mass. The first of the
$\alpha_{s}^2$ terms give the Coulombic corrections ($\rm C$) the second
ones the non-Coulombic ones ($\rm nC$). Here and below we use
$\alpha_{s}=\alpha_{s}(\mu)$ as defined in the $\overline{\mbox{MS}}$
scheme and the $\mu$ dependence from the numerical solution of the three
loop renormalization group equation with the starting value
$\alpha_{s}(M_{Z})= 0.1176$. We use $n_{f}=5$. This is well justified,
as long as the characteristic scale $\alpha_{s}m_{\tilde{g}}$ is smaller
than the top-quark mass, an assumption well justified for all gluino masses
under consideration. Top-quark effects could be included following
\cite{Melles:1998dj,Melles:2000dq}. If not stated otherwise, in the
following we use for the renormalization scale
$\mu=\mu_S$ with $\mu_S=m_{\tilde{g}}C_{A}\alpha_{s}(\mu_S)/n$ 
corresponding to $L_{n}=0$.

We have convinced ourselves that the perturbative series and the
numerical results are in very good agreement as far as $V_{\rm C}$ is
concerned. The subsequent analysis will therefore be based on the fully
perturbative approach.

\begin{figure}[t]
\begin{center}
\begin{minipage}{13 cm}
\includegraphics[angle=270,width=\textwidth]{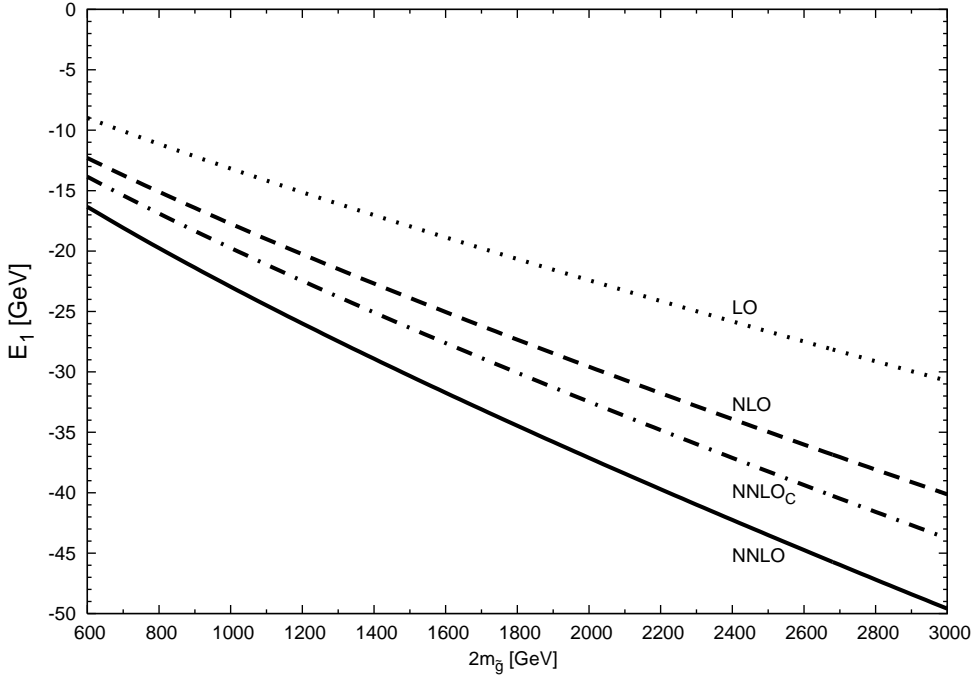}
\caption{Ground state energy $E_{1}$ in the pole mass scheme as function
of twice the constituent mass.}
\label{EnergiePol}
\end{minipage}
\end{center}
\end{figure}

\begin{figure}[t]
\begin{center}
\begin{minipage}{13 cm}
\includegraphics[angle=270,width=\textwidth]{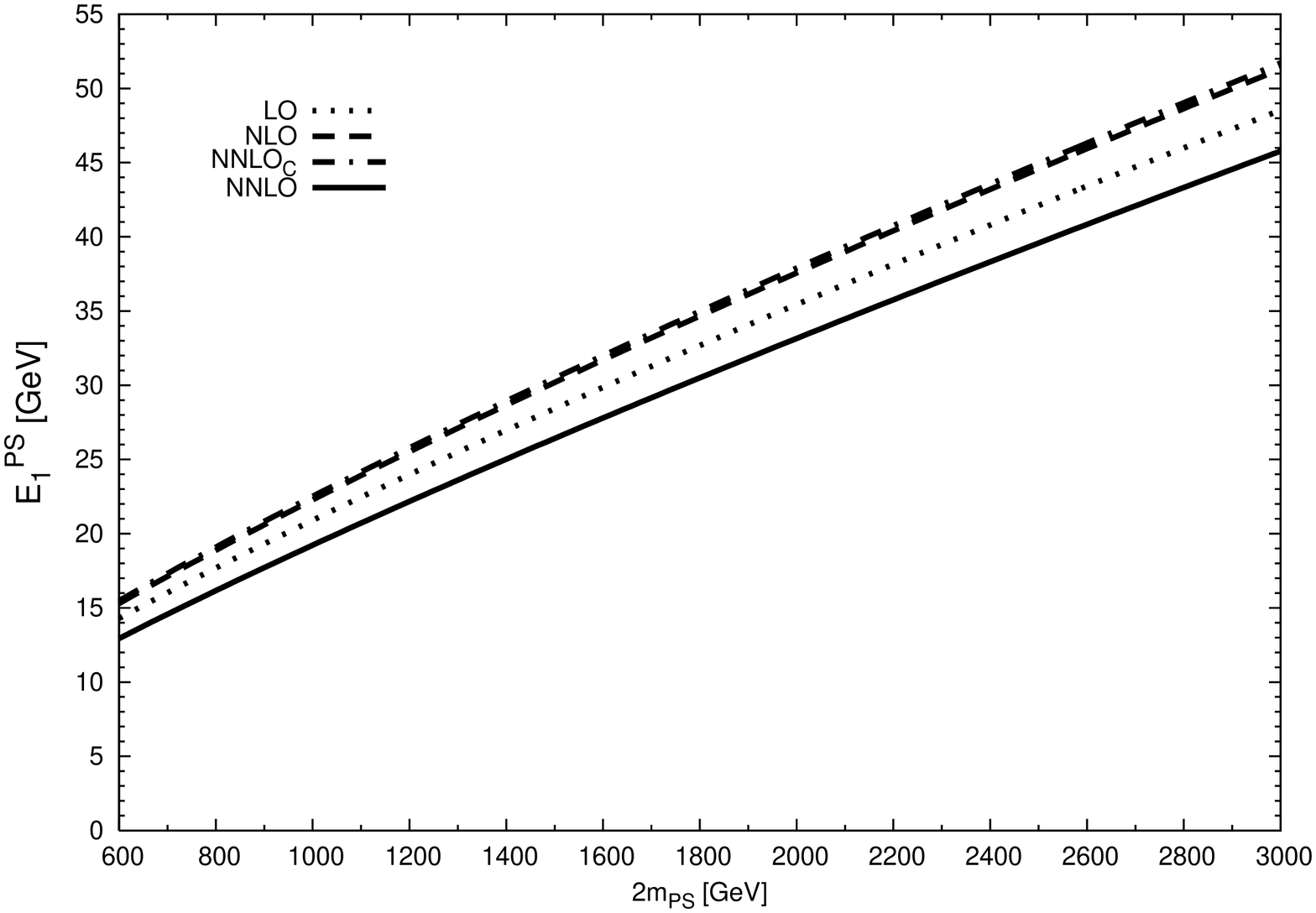}
\caption{Ground state energy $E_{1}$ in the potential subtracted mass
  scheme as function of twice the potential subtracted mass. The
  curves have been obtained using the pole mass as input and evaluating
  both the potential subtracted mass (using Eq.~(\ref{mPS})) and
  $E_1^{PS}$ to a given order in
  $\alpha_s$. Note that the dash-dotted and dashed curves are almost on top of
  each other.}
\label{EnergiePS}
\end{minipage}
\end{center}
\end{figure}

The groundstate energy $E_{1}$ is shown in Fig.~\ref{EnergiePol} as a
function of $2m_{\tilde{g}}$ for gluino masses $m_{\tilde{g}}$ between
$300\,\mbox{GeV}$ and $1500\,\mbox{GeV}$. The LO (dotted), NLO (dashed)
and NNLO predictions are displayed individually. For the NNLO result
both the result for $V_{\rm C}$ (dash-dotted) and the one including the
corrections collected in $U_{1}$, $U_{2}$ and $U_{3}$ (solid) are shown.
As stated above, $m_{\tilde{g}}$ is understood as the constituent pole
mass. The poor convergence of the perturbative series can partly be
traced back to the large non-Coulombic correction which appear for the
first time in NNLO and amount to twice the NNLO Coulombic one.
Qualitatively this behavior is quite similar to the one observed for
quarkonia in \cite{Beneke:2005hg,Penin:2005eu,Penin:2002zv}. It remains
to be seen to which extent inclusion of orders of $\mbox{N}^3\mbox{LO}$
and higher will stabilize these predictions.\footnote{For the quarkonium
  case, see Refs.~\cite{Penin:2002zv,Kiyo:2002rr}.}
In Fig.~\ref{EnergiePS} we show as an alternative representation the
predictions for the ground state energy, using as reference the
potential subtracted mass \cite{Beneke:1998rk}, which is related to the
pole mass through
\begin{eqnarray}
m_{\tilde{g}}-m_{\rm PS}\left(\mu_{f}\right)&=&\delta
m\left(\mu_{f}\right)\,\,=\,\,-\frac{1}{2}\int_{\left|\vec{q}\,\right|<\mu_{f}}\frac{d^3
  q}{\left(2\pi\right)^3}\tilde{V}\left(q\right)
\nonumber\\
&=&\frac{C_{A}\alpha_{s}(\mu)}{\pi}\mu_{f}\biggl\{1+\frac{\alpha_{s}(\mu)}{4\pi}\biggl[a_{1}-\beta_{0}\left(\ln\left(\frac{\mu_{f}^2}{\mu^2}\right)-2\right)\biggr]\biggr.
\nonumber\\
&&
\biggl.\mbox{}+\left(\frac{\alpha_{s}(\mu)}{4\pi}\right)^2\biggl[a_{2}-\left(2a_{1}\beta_{0}+\beta_{1}\right)\left(\ln\left(\frac{\mu_{f}^2}{\mu^2}\right)-2\right)\biggr.\biggr.
\nonumber\\
&&
\biggl.\biggl.\mbox{}+\beta_{0}^2\left(\ln^2\left(\frac{\mu_{f}^2}{\mu^2}\right)-4\ln\left(\frac{\mu_{f}^2}{\mu^2}\right)+8\right)\biggr]\biggr\}
\,.
\label{mPS}
\end{eqnarray}
In the following we adopt for the factorization scale
$\mu_{f}=m_{\tilde{g}}C_{A}\alpha_{s}(\mu_S)$, independent of $n$
(with $\mu_S=m_{\tilde{g}}C_{A}\alpha_{s}(\mu_S)/n$ as before).
(Note, that $m_{\rm PS}$ now also depends on $n$.)
We refrain from listing explicitly the results for the energy levels in the
potential subtracted scheme since the corresponding formulae 
are quite bulky. They can easily be obtained from Eqs.~(\ref{EundPsiNum})
and~(\ref{mPS}). 

\begin{figure}[t]
\begin{center}
\begin{minipage}{13 cm}
\includegraphics[angle=270,width=\textwidth]{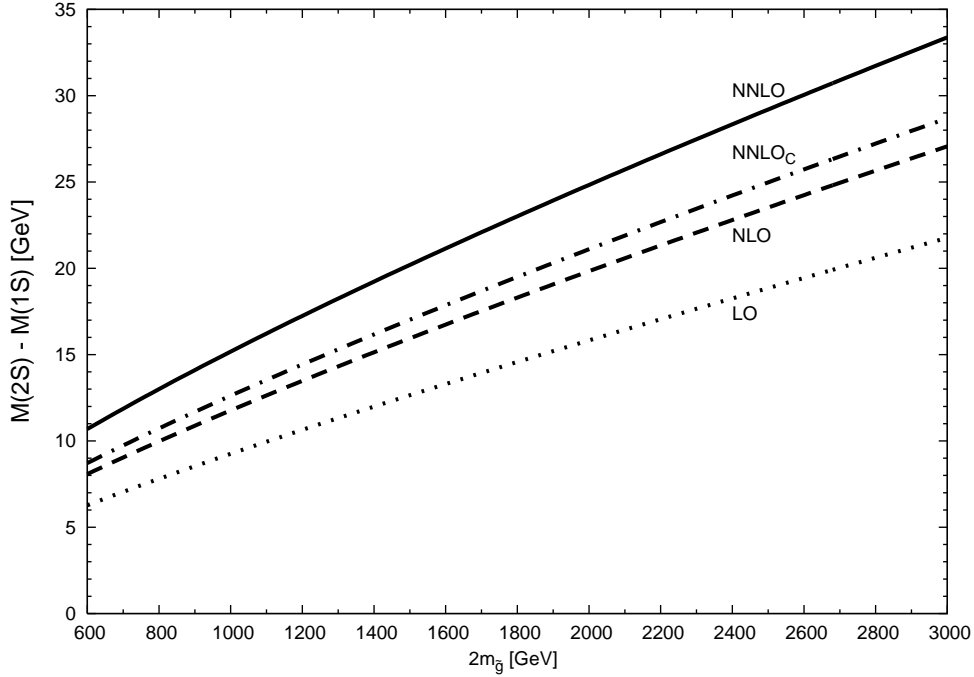}
\caption{Meson mass difference $M(2S)-M(1S)$ as
  function of twice the constituent mass.}
\label{AnregungPol}
\end{minipage}
\end{center}
\end{figure}

The implementation of the potential subtracted mass
leads to a significantly improved convergence of
the perturbative series for $E_{1}^{\rm PS}$ as far as the Coulombic
part is concerned (see Fig.~\ref{EnergiePS}) and similarly for
$E_{2}^{\rm PS}$. However, the meson mass difference between the $1S$
and the $2S$ state, $M(2S)-M(1S)$, is independent of the choice of the scheme
and exhibits a poor convergence in both cases.
The predictions for $M(2S)-M(1S)$ are shown in Fig.~\ref{AnregungPol},
adopting $\mu=\mu_S$. The NNLO terms are evidently
important and the poor convergence of the prediction for $E_{1}$ is
reflected in this figure.

In the on-shell scheme the $1S$ binding energy is evidently closely
related to the energy difference between the groundstate and the onset
of open gluino production or, more specifically, the threshold for pair
production of color-neutral $\left(\tilde{g}g\right)$ hadrons. However,
anticipating a mass resolution of several tens of GeV at least, it will
be difficult to resolve the densely distributed radial excitations. For
this reason the energy difference between the groundstate and the first
radial excitation is a convenient measure of the isolation of the
$1S$ state.

\begin{figure}[t]
\begin{center}
\begin{minipage}{13 cm}
\includegraphics[angle=270,width=\textwidth]{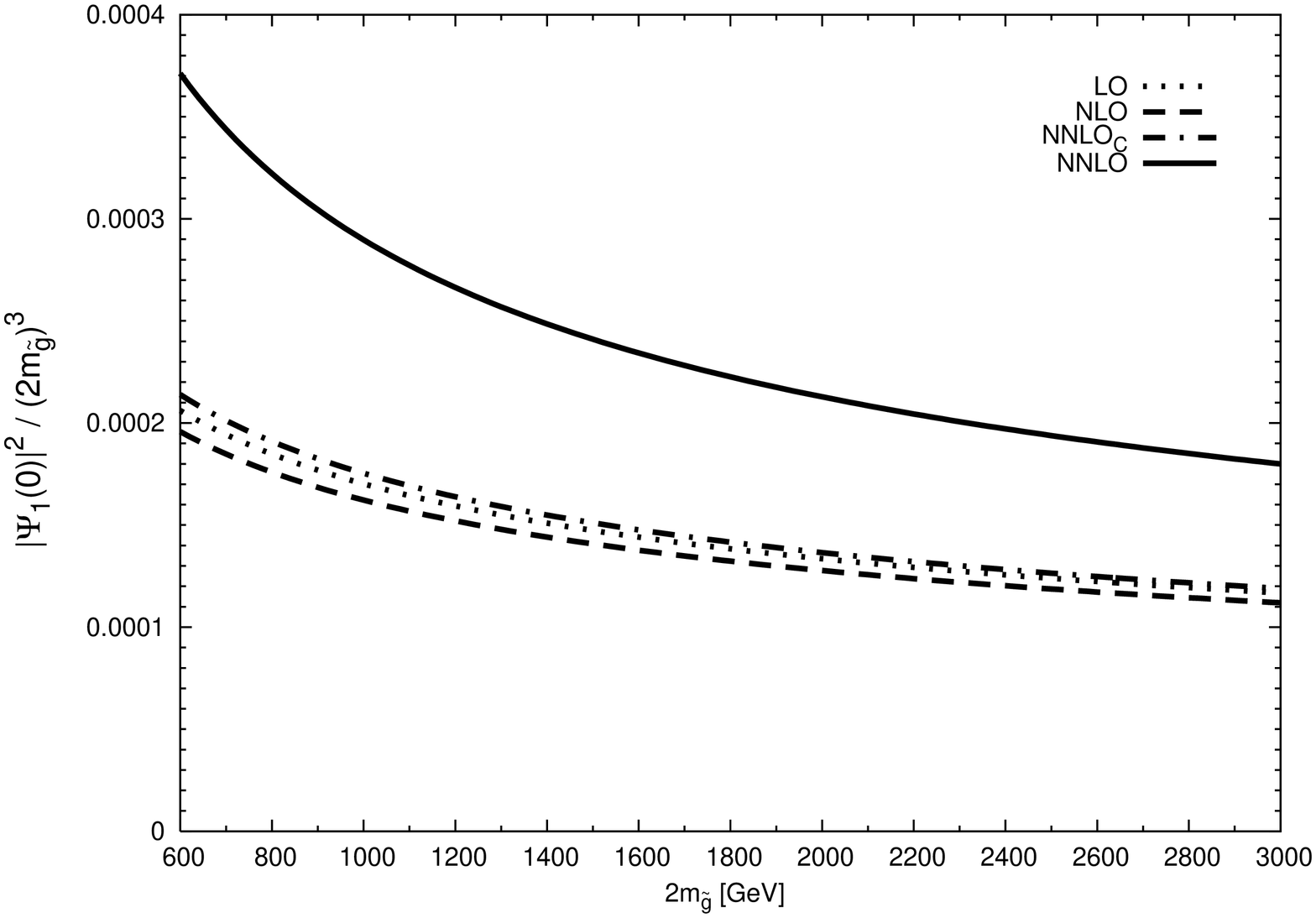}
\caption{Squared ground state wave function at the origin as
  function of twice the constituent mass.}
\label{Wellenfunktion}
\end{minipage}
\end{center}
\end{figure}

Production and decay of non-relativistic $S$-wave bound states are proportional
to the square of the wave function at the origin. The analytic results
are collected in the Appendix~\ref{app:energy}, numerical results are listed in
Eq.~(\ref{EundPsiNum}) and shown in Fig.~\ref{Wellenfunktion}. The
enormous size in particular of the non-Coulombic NNLO terms has also
been observed for the top-antitop system and destabilizes the
predictions \cite{Hoang:2000yr}. At present the size of these
corrections must be considered as an estimate of the theory
uncertainty.

\begin{figure}[t]
\begin{center}
\includegraphics[angle=270,width=.8\textwidth]{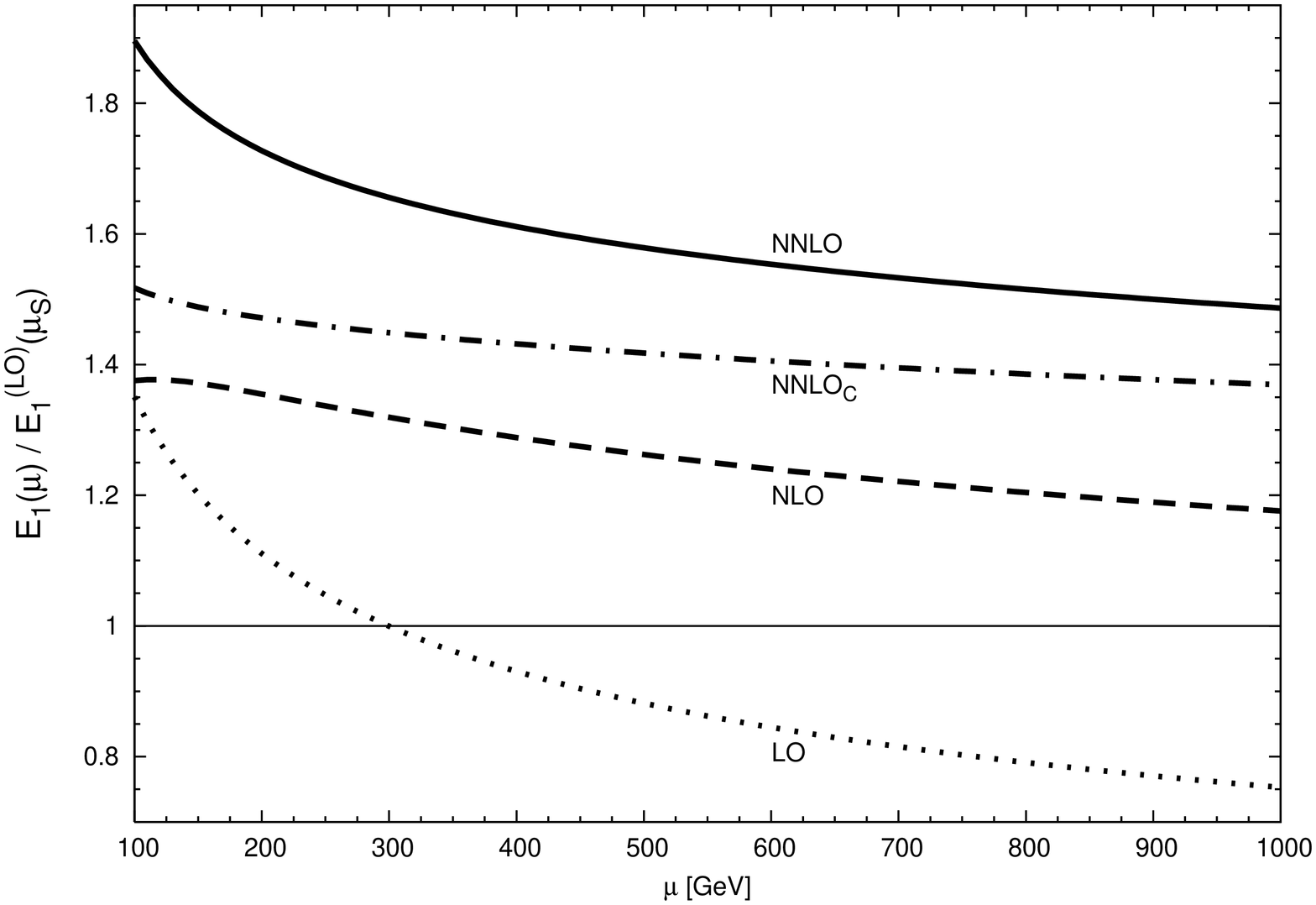}
\includegraphics[angle=270,width=.8\textwidth]{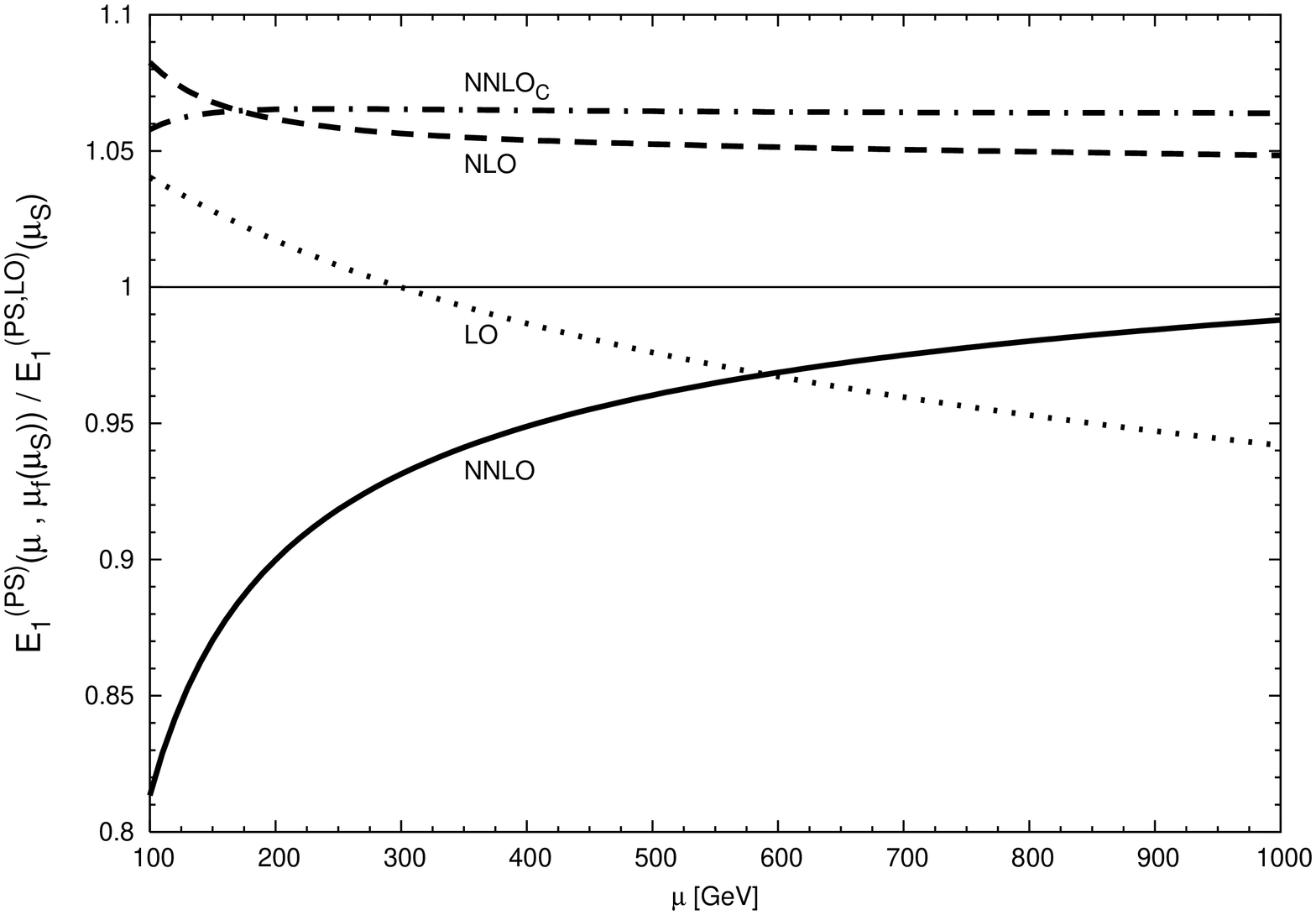}
\caption{Renormalization scale dependence of the ground state
  energy for $m_{\tilde{g}}=1\,\mbox{TeV}$ in the pole mass scheme (a) and the
  potential subtracted scheme (b).}
\label{EnergieMU}
\end{center}
\end{figure}

\begin{figure}[t]
\begin{center}
\begin{minipage}{13 cm}
\includegraphics[angle=270,width=\textwidth]{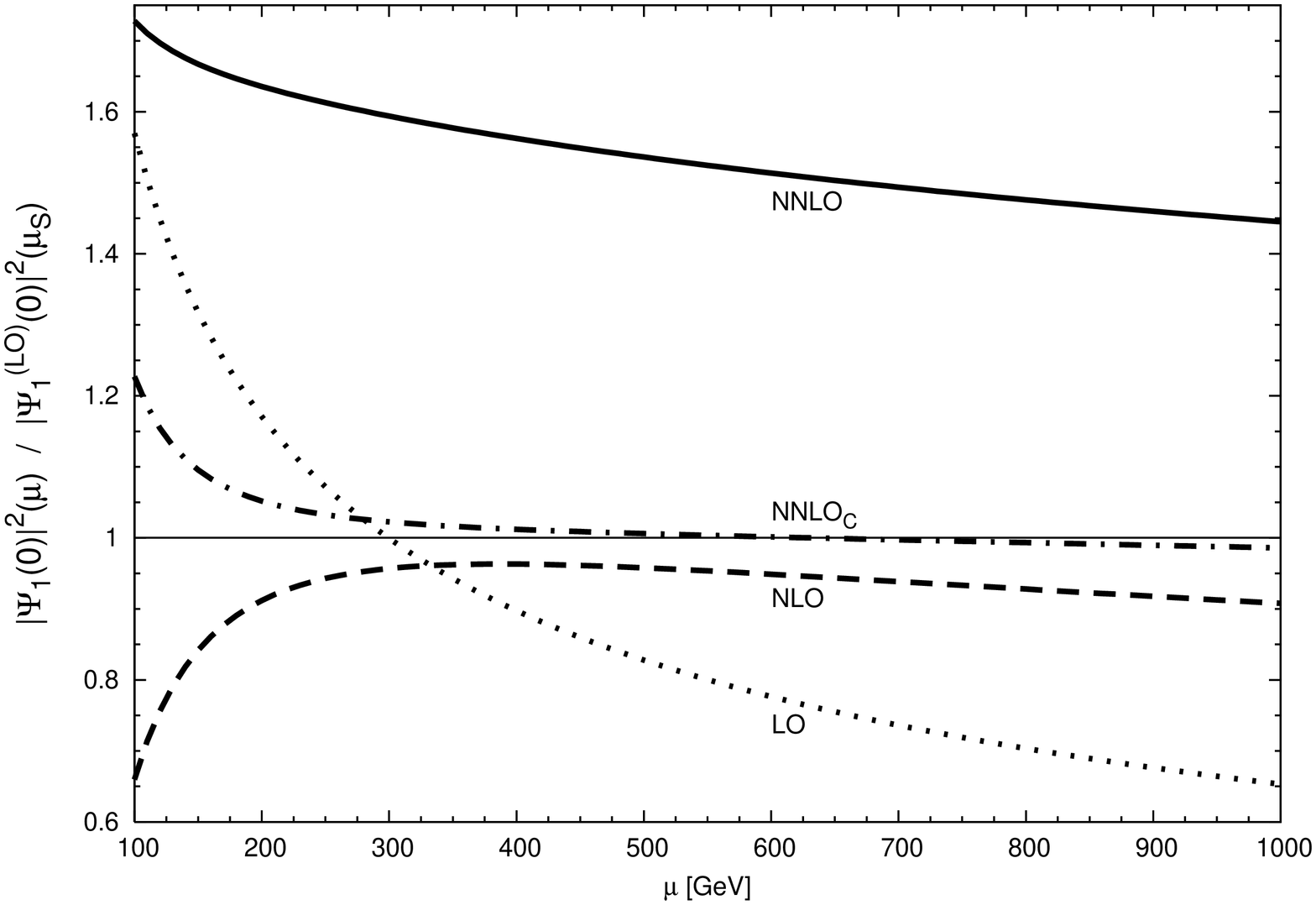}
\caption{Renormalization scale dependence of the squared ground state
  wave function at the origin for $m_{\tilde{g}}=1\,\mbox{TeV}$.}
\label{WellenfunktionMU}
\end{minipage}
\end{center}
\end{figure}

The renormalization scale dependence of the LO, NLO and
$\mbox{NNLO}_{C}$ results for $E_{1}$ and
$\left|\Psi_{1}\left(0\right)\right|^2$ are shown in
Figs.~\ref{EnergieMU} and \ref{WellenfunktionMU} for a
constituent mass of $1$ TeV. The predictions are normalized to the LO
prediction evaluated at the scale $\mu$ as defined above. A
significant stabilization is observed with increasing order in
$\alpha_{s}$ and, furthermore, the higher order
contributions are reasonably small for the preferred choice of the
renormalization scale, at least, as far as the Coulombic part is
concerned. {This is in particular true for the ground state energy
parameterized in terms of the potential subtracted mass where close to
$\mu=150$~GeV the Coulombic NNLO results even vanish.
In the case of the wave function one observes around $\mu=300$~GeV
corrections of the order of a few percent from NLO and NNLO$_{\rm C}$,
however, a huge contribution from the non-Coulombic terms.}


\section{\label{sec:production}Bound state production and decay}

In the present context squarks are assumed to be significantly heavier
than gluinos, such that the direct decay
$\tilde{g}\rightarrow\tilde{q}g$ is forbidden. An extreme example is
provided by Split SUSY \cite{ArkaniHamed:2004fb,Giudice:2004tc} with
$m_{\tilde{q}}\gg m_{\tilde{g}}$ and we shall adopt this simplifying
mass assignment throughout this section. However, this assumption could
be significantly relaxed. Indeed, once $m_{\tilde{q}}>m_{\tilde{g}}$ is
valid the gluinos have a very small decay rate, for $m_{\tilde{q}} \gg
m_{\tilde{g}}$ of order \cite{Haber:1984rc}
\begin{eqnarray}
\Gamma\left(\tilde{g}\rightarrow\tilde{\gamma}q\overline{q}\right)&\approx&\left(\sum
  _{q}e_{q}^2\right)\frac{\alpha\alpha_{s}}{48\pi}\frac{m_{\tilde{g}}^5}{m_{\tilde{q}}^4}
\nonumber\\
&\approx&10\,\mbox{MeV}\left[\frac{m_{\tilde{g}}^5}{m_{\tilde{q}}^4\,\mbox{TeV}}\right]
\,,
\label{singleDecay}
\end{eqnarray}
characterizing the rate for gluino decays into neutralinos
(similar estimates being applicable for decays into
charginos)
such that the decay rate is far
smaller than the annihilation decay.

The evaluation of the rate for gluinonium annihilation into gluon (and
quark) jets proceeds similar to the one for quarkonia. In Born
approximation the color factor is replaced as follows
\begin{eqnarray}
\sum_{a,b}\left|\frac{\delta_{ik}}{\sqrt{N_{C}}}\frac{\lambda^{a}_{ij}}{2}\frac{\lambda^{b}_{jk}}{2}\right|^2&=&\frac{2}{3}\hspace{0.7cm}\Rightarrow\hspace{0.7cm}\sum_{a,b}\left|\frac{\delta_{ik}}{\sqrt{N_{C}^2-1}}f^{aij}f^{bjk}\right|^2=9\,
\label{colorReplace}
\end{eqnarray}
and another relative factor $\frac{1}{2}$ arises from the Majorana
nature of the gluinos, whence
\cite{Keung:1983wz,Kuhn:1983sc,Goldman:1984mj}
\begin{eqnarray}
  \Gamma_{\rm LO}\left(0^{-+}\rightarrow gg\right)&=&\Gamma_{gg}
  \,\,=\,\,\frac{C_{A}^2}{2}\frac{\alpha_{s}^2}{m_{\tilde{g}}^2}\left|R(0)\right|^2\,,
  \label{GammaQQ}
\end{eqnarray}
with $|R(0)|^2=4\pi|\Psi_1(0)|^2$.
Using the wave function obtained in lowest order perturbation theory
leads to $\Gamma_{gg}\approx(C_{A}\alpha_{s})^5m_{\tilde{g}}/4$ which
provides a qualitative estimate of the full result. The perturbative
corrections arising from virtual and real emission can be calculated
similar to the ones for quarkonia \cite{Kuhn:1992qw}. They reduce the
scale dependence of the lowest order approximation and are complementary
to the higher order corrections to the wave function. In total one finds
\begin{eqnarray}
\Gamma_{\rm
  NLO}&=&\Gamma_{gg}\Biggl\{1+\frac{\alpha_{s}(\mu)}{\pi}
\biggl[C_{A}\left(\frac{109}{18}-\frac{7}{24}\pi^2\right)-\frac{16}{9}n_{f}T_{F}\biggr.\Biggr.
\nonumber\\
&&\mbox{}
\Biggl.\biggl.+\left(\frac{11}{6}C_{A}-\frac{2}{3}n_{f}T_{F}\right)\ln\left(\frac{\mu^2}{4m_{\tilde{g}}^2}\right)\biggr]\Biggr\}
\,.
\nonumber\\
\label{GammaNLO}
\end{eqnarray}

\begin{figure}[t]
\begin{center}
\begin{minipage}{13 cm}
\includegraphics[angle=270,width=\textwidth]{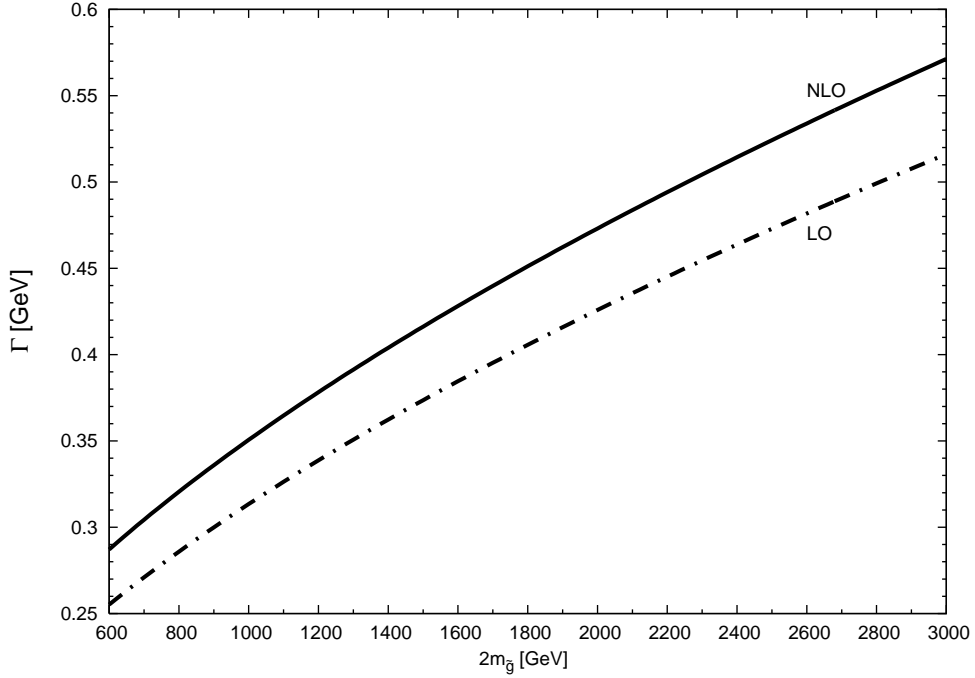}
\caption{Decay rate of the $1S$ state into gluons.}
\label{Zerfall}
\end{minipage}
\end{center}
\end{figure}

Here $\alpha_{s}$ is understood in the theory with $n_{f}=6$ effective
flavors and all quark masses, including $m_{t}$, have been neglected.
The $\alpha_{s}^3$ term present in $\left|R(0)\right|^2$ is interpreted
in the $n_{f}=5$ effective theory. The predictions for $\Gamma_{\rm LO}$
and $\Gamma_{\rm NLO}$ are shown in Fig.~\ref{Zerfall}, where the NNLO
wave function as shown in Fig.~\ref{Wellenfunktion} (excluding the
non-Coulombic contribution) and $\mu=2m_{\tilde{g}}$ for the
renormalization scale in Eq.~(\ref{GammaNLO}) have been adopted. As
anticipated, the decay rate of the bound state is small compared to the
level spacing and large compared to
$\Gamma(\tilde{g}\rightarrow\tilde{\gamma}q\overline{q})$, the decay
rate of a single gluino, once $m_{\tilde{g}}<m_{\tilde{q}}$.  Hence
annihilation decays will constitute the dominant signal. However, it
should be emphasized, that bound states will exist even for
$m_{\tilde{g}}>m_{\tilde{q}}$, as long as
$\Gamma\left(\tilde{g}\rightarrow\tilde{q}q\right)\leq E_{2}-E_{1}$,
which is true for the region $m_{\tilde{q}}\gsim0.9\,m_{\tilde{g}}$.

\begin{figure}[t]
\begin{center}
\begin{minipage}{13 cm}
\includegraphics[angle=270,width=\textwidth]{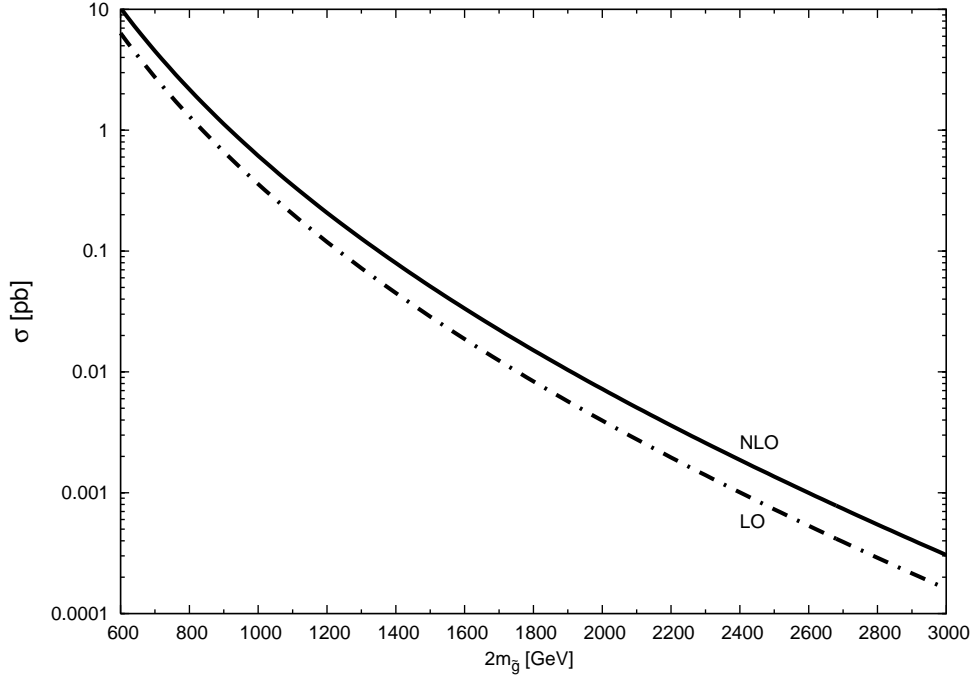}
\caption{LO and NLO production cross section for the $1S$ state at
  $\sqrt{s}=14~\mbox{TeV}$.}
\label{Produktion}
\end{minipage}
\end{center}
\end{figure}

The bound state production cross section can be calculated similarly to
the one for the pseudoscalar bound state of two top quarks (usually
called $\eta_{t}$) \cite{Kuhn:1992qw}. In Born approximation the
reaction proceeds through gluon fusion and the radiative corrections
involve additional $gq$, $g\overline{q}$ and $q\overline{q}$ initiated
subprocesses. The final result is quite similar to the one for quarkonia
production \cite{Kuhn:1992qw}. We find
\begin{eqnarray}
\hspace{-1cm}
\sigma_{\rm had}(S)&=&\sum_{ab}\int_{0}^{1}{\rm d}x_{1}
\int_{0}^{1}{\rm d}x_{2}f_{a}^{h_{1}}(x_{1},\mu_{F}^2)f_{b}^{h_{2}}(x_{2},\mu_{F}^2)
\hat{\sigma}_{ab}(s=x_{1}x_{2}S)
\,,\nonumber\\
\hat{\sigma}_{gg}&=&\sigma_{0}\biggl\{\delta(1-z)+\frac{\alpha_{s}}{\pi}
\biggl[-P_{gg}(z)\ln\left(\frac{\mu^2}{4m_{\tilde{g}}^2}\right)+C_{A}F(z)\biggr.\biggr.
\nonumber\\
&&
\biggl.\biggl.\mbox{}
+\delta(1-z)\left(\left(\frac{11}{6}C_{A}-\frac{2}{3}n_{f}T_{F}\right)\ln\left(\frac{\mu^2}{4m_{\tilde{g}}^2}\right)+C_{A}\left(-4+\frac{1}{3}\pi^2\right)\right)\biggr]\biggr\}
\,,\nonumber\\
\hat{\sigma}_{gq}&=&\sigma_{0}\frac{\alpha_{s}}{\pi}\biggl\{-C_{F}\frac{z}{2}\ln
z+C_{F}z+\frac{1}{2}P_{gq}(z)\left(\ln\left(\frac{4m_{\tilde{g}}^2(1-z)^2}{\mu^2}\right)-1\right)\biggr\}
\,,\nonumber\\
\hat{\sigma}_{g\overline{q}}&=&\hat{\sigma}_{gq}
\,,\nonumber\\
\hat{\sigma}_{q\overline{q}}&=&\sigma_{0}\frac{\alpha_{s}}{\pi}\frac{32}{27}z(1-z)
\,,
\end{eqnarray}
with
\begin{eqnarray}
\sigma_{0}&=&\frac{C_{A}^2\pi^2\alpha_{s}^2}{4}\frac{|R(0)|^2}{s(2m_{\tilde{g}})^3}
\,,\nonumber\\
P_{gg}(z)&=&2C_{A}\left(\frac{1}{z}+\left[\frac{1}{1-z}\right]_{+}+z(1-z)-2\right)
\nonumber\\
&&+\left(\frac{11}{6}C_{A}-\frac{2}{3}n_{f}T_{F}\right)\delta(1-z)
\,,\nonumber\\
P_{gq}(z)&=& C_{F}\frac{1+(1-z)^2}{z}
\,,\nonumber\\
F(z)&=&\frac{11z^5+11z^4+13z^3+19z^2+6z-12}{6z(1+z)^2}\biggr.
\nonumber\\
&&\mbox{}+\,4\left(\frac{1}{z}+z(1-z)-2\right)\ln(1-z)+4\left[\frac{\ln(1-z)}{1-z}\right]_{+}
\nonumber\\
&&\mbox{}\biggl.+\left(\frac{2(z^3-2z^2-3z-2)(z^3-z+2)z\ln z}{(1+z)^3(1-z)}-3\right)\frac{1}{1-z}
\,.\label{sigmaADD}
\end{eqnarray}
Here $z=4m_{\tilde{g}}^2/s\le1$, $\mu$ and $\mu_{F}$ are the renormalization
and the factorization scales, $\alpha_{s}$ is again defined in the
$n_{f}=6$ flavor theory, $P_{ij}$ are the Altarelli-Parisi splitting
functions and $[\ldots]_{+}$ denotes the usual $+$ distribution. For the
wave function at the origin we again adopt the NNLO result without the
non-Coulombic terms. The predictions for the production cross section
are shown in Fig.~\ref{Produktion} where the parton distribution functions 
MSTW2008LO (MSTW2008NLO)~\cite{Martin:2009iq} have been used for the LO (NLO)
calculation. 
Leading order (dashed) and NLO (solid curve) predictions are
evaluated using $\mu=\mu_{F}=2m_{\tilde{g}}$.

\begin{figure}[htb]
\begin{center}
\begin{minipage}{13 cm}
\includegraphics[angle=270,width=\textwidth]{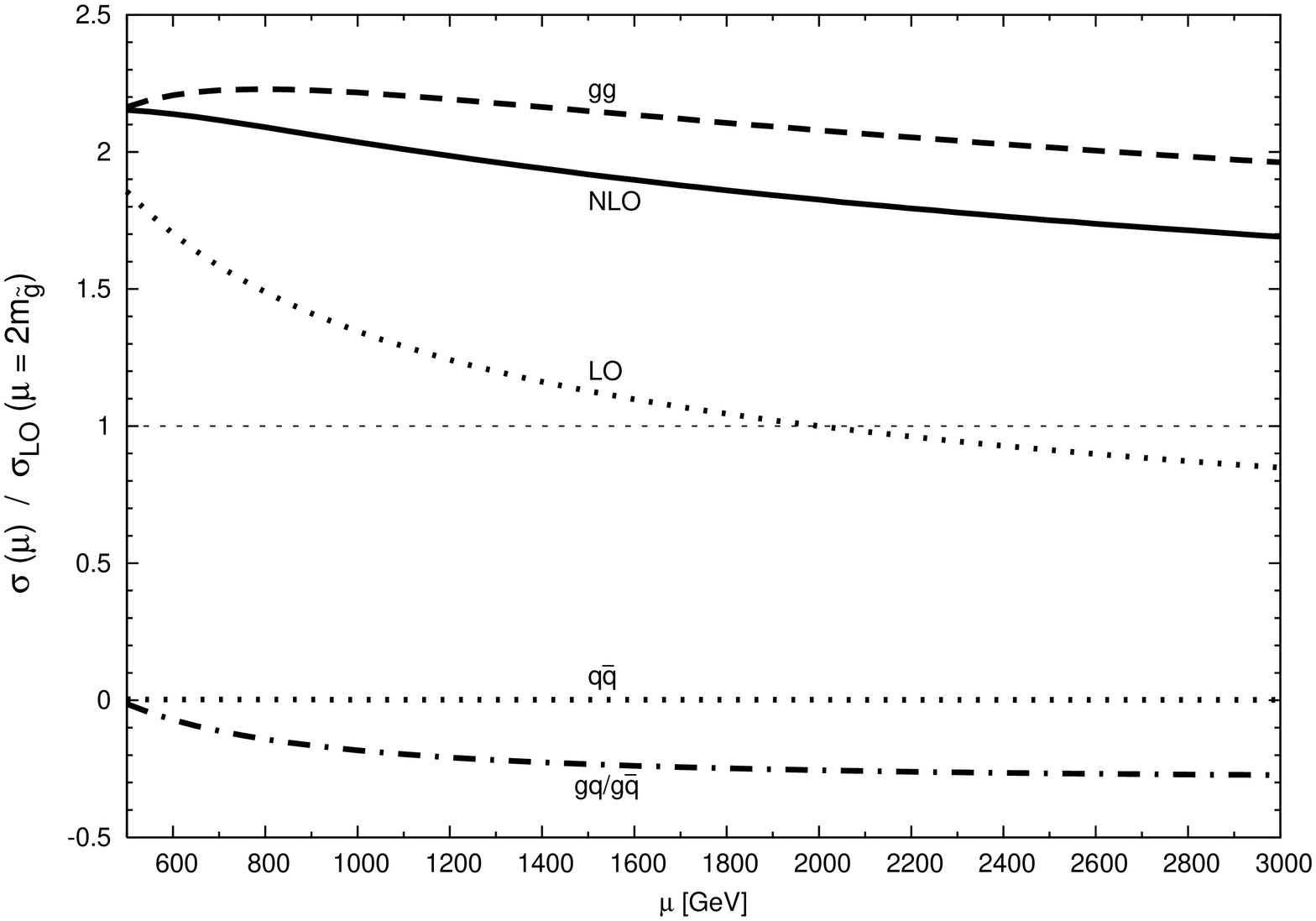}
\caption{Renormalization- and factorization-scale dependence of the $1S$
  production cross section and decomposition into contributions from
  various subprocesses. The line labeled ``gg'' contains also the LO result.}
\label{ProduktionMU}
\end{minipage}
\end{center}
\end{figure}

For a gluino mass of $1~\mbox{TeV}$ the $\mu$ dependence (with $\mu_F=\mu$) 
of the LO and
the NLO results is shown in Fig.~\ref{ProduktionMU}, together with the
decomposition of the NLO result into the contributions from the
different subprocesses.

From Figs.~\ref{Produktion} and \ref{ProduktionMU} we learn that the NLO
correction amounts to more than $+70\%$ fairly independent of the
renormalization scale. The dominant uncertainty, not displayed in these
figures, arises from the wave function at the origin, with indications
for an increase by another factor of two.

For an anticipated integrated luminosity of $100~\mbox{fb}^{-1}$ between
$10$ and $10^6$ events are expected for $m_{\tilde{g}}$ between
$1.5~\mbox{TeV}$ and $300~\mbox{GeV}$. It remains to be seen, if these can be 
separated from the hadronic continuum background.

\section{\label{sec:photons}Annihilation decays into top quarks and
  into two photons}
In the limit of extremely heavy squarks (and conserved R-parity),
corresponding to Split SUSY, decay modes of single gluinos are extremely
suppressed, such that gluinos (confined in color-neutral hadrons
like $\tilde{g}g$) could even travel microscopic distances. In this case
gluinonium annihilation into two gluons is the only relevant
channel. However, once squark and gluino masses are comparable, new
annihilation decays become possible. At first glance one might consider
the mode $\left(\tilde{g}\tilde{g}\right)\rightarrow q\overline{q}$,
mediated by squark exchange. However, for massless quarks this process
vanishes as a consequence of helicity conservation. For massive
top quarks one finds
\begin{eqnarray}
\hspace{-0.5cm}
R_{t\overline{t}} &=& 
\frac{\Gamma\left(0^{-+}\rightarrow
    t\overline{t}\right)}{\Gamma\left(0^{-+}\rightarrow gg\right)} \,\,=\,\,
\frac{4C_{F}}{C_{A}^2}\sqrt{1-\frac{m_{t}^2}{m_{\tilde{g}}^2}}\frac{m_{\tilde{g}}^2m_{t}^2}{\left(m_{\tilde{g}}^2+m_{\tilde{t}}^2-m_{t}^2\right)^2}
\,.\label{rTT}
\end{eqnarray}
using $m_{\tilde{t}}=m_{\tilde{t}_{1}}=m_{\tilde{t}_{2}}$. The
complete result for $m_{\tilde{t}_{1}}\neq m_{\tilde{t}_{2}}$ and
including squark mixing is given in Appendix~\ref{app:ttbar}.

\begin{figure}[t]
\begin{center}
\begin{minipage}{13 cm}
\includegraphics[angle=270,width=\textwidth]{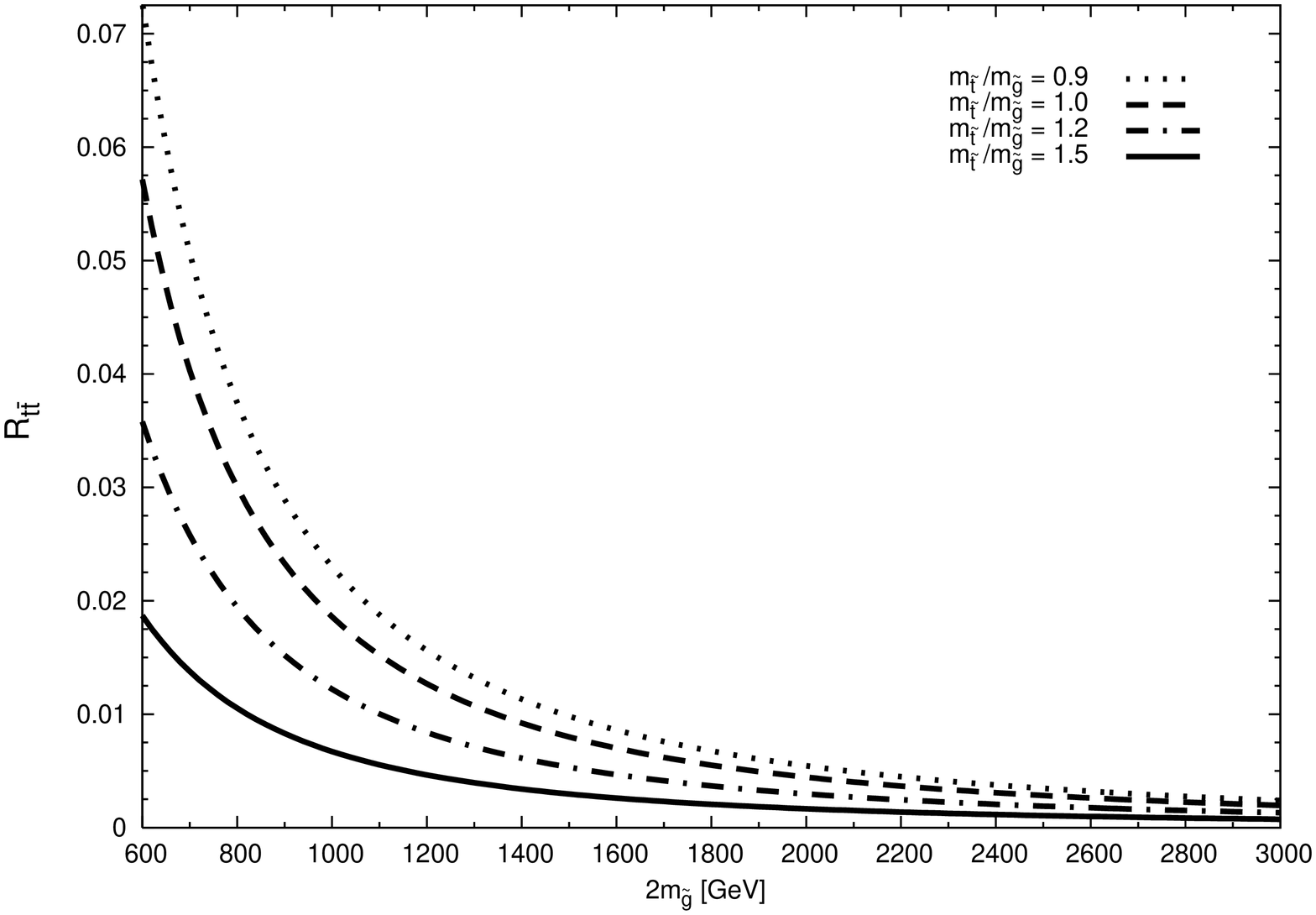}
\caption{Branching ratio of the $1S$ decay rate into top quarks.}
\label{RatioTT}
\end{minipage}
\end{center}
\end{figure}

The result is plotted in Fig.~\ref{RatioTT} for different ratios of
$m_{\tilde{t}}/m_{\tilde{g}}$. Branching ratios of up to $7\%$ are
possible for gluino masses close to $300\ \mbox{GeV}$ and squark masses
not much bigger. For heavier squark and gluino masses the ratio drops
quickly.

As a second possibility we consider the loop induced decay into two
photons. Although suppressed by the electromagnetic couplings
$\alpha^{2}$, the cleaner signal-to-background ratio might (at least in
principle) compensate this disadvantage. For arbitrary squark mixing and
masses the result is given in Appendix~\ref{app:gamgam}. In the limit of
massless quarks and degenerate squarks the decay rate simplifies
considerably
\begin{eqnarray}
  R_{\gamma\gamma}&=&\frac{\Gamma\left(0^{-+}\rightarrow\gamma\gamma\right)}{\Gamma\left(0^{-+}\rightarrow
      gg\right)} \,\,=\,\,
  \frac{4T_{F}}{C_{A}^2}\frac{\alpha^2}{\pi^2}\left(\sum_{f}Q_{f}^2\right)^2\left|\mbox{Li}_{2}\left(-\frac{m_{\tilde{g}}^2}{m_{\tilde{q}}^2}\right)-\mbox{Li}_{2}\left(\frac{m_{\tilde{g}}^2}{m_{\tilde{q}}^2}\right)\right|^2
\label{RatioGamGam}
\,.
\end{eqnarray}
With a ratio $R_{\gamma\gamma}$ around one to several $10^{-5}$ the
two-photon signal will not be detectable.


\section{\label{sec:background}Signal versus background}

Assuming the two-gluon mode to dominate the decay of gluino bound
states, the resonant signal must be isolated from the two-jet continuum.
Below we present an order of magnitude estimate for the signal-to-background
ratio, following essentially the arguments given in
Ref.~\cite{Goldman:1984mj}. For simplicity it is assumed that quark and gluon 
jets can be experimentally separated. In this case the background is
dominated by gluon-gluon scattering. Furthermore, we enhance the
signal-to-background ratio by excluding small scattering angles in the 
partonic center-of-mass system, requiring $\left|\cos(\theta)\right|\leq z<1$.

To isolate a resonance in the gluon-gluon channel, a good experimental
resolution $\Delta M$ of the invariant dijet mass $M_{jj}$ is
essential. Let us compare the differential cross section, integrated
over the interval $M-\frac{\Delta M}{2}\leq M_{jj}\leq M+\frac{\Delta
  M}{2}$ for background and signal, and integrated over the same angular
range. This leads to the signal-to-background ratio
\begin{eqnarray}
\hspace{-0.8cm}r(z,\Delta M)&=&\frac{S(z,\Delta M)}{B(z,\Delta M)}=\frac{\pi}{3}\frac{\Gamma_{gg}}{\Delta M\alpha_{s}^2}\left[\frac{129-32z^2-z^4}{1-z^2}-\frac{24}{z}\log\left(\frac{1+z}{1-z}\right)\right]^{-1}
.\label{soverr}
\end{eqnarray}

\begin{figure}[htb]
\begin{center}
\begin{minipage}{13 cm}
\includegraphics[angle=270,width=\textwidth]{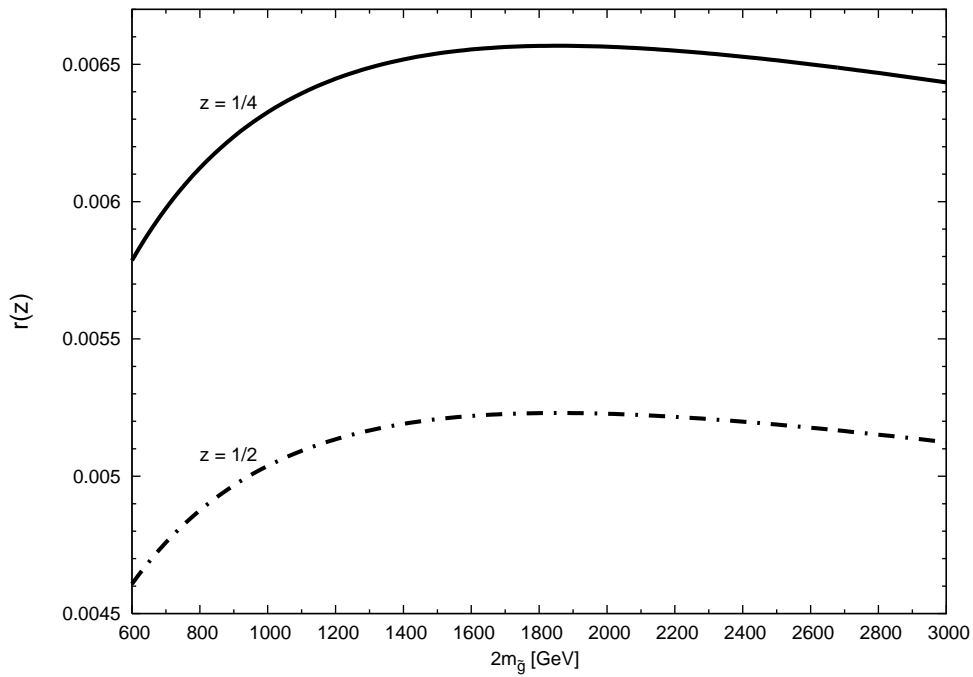}
\caption{Signal-to-background ratio for different values of $z$.}
\label{Back}
\end{minipage}
\end{center}
\end{figure}

Using $\Delta M/M_{jj} = 0.038 + 38/M_{jj}$ for the dijet resolution
\cite{cms:2007}, one obtains the signal-to-background ratio as shown in
Fig.~\ref{Back}, varying between $0.4\%$ to $0.7\%$. Given event rates
between $10^3$ and $10^5$ for $m_{\tilde{g}}$ between $500\ \mbox{GeV}$
and $1\ \mbox{TeV}$ the detection of gluino bound states might become
feasible, in particular at a high luminosity version of the LHC.


\section{\label{sec:conclusions}Conclusions}
For a large class of supersymmetric extensions of the Standard Model gluinos
will be sufficiently stable to form color-neutral non-relativistic
boundstates. The NNLO potential between color-octet constituents forming
color singlets has been derived. Just as in leading and next-to-leading
order the Coulombic part of the NNLO potential is related to the
quarkonium potential through the substitution $C_{F}\rightarrow C_{A}$
in the overall factor. The transformation of the short distance terms
involving an additional factor $1/m$ is only slightly more complicated
and discussed in Section~\ref{sec:properties}. Based on this
potential, predictions for the binding energy and the wave function at
the origin of the lowest-lying boundstates and the first radial
excitation have been derived. For gluino masses between $300\ \mbox{GeV}$
and $1500\ \mbox{GeV}$ the $1S$ binding energy (in the pole mass scheme)
varies between $15\ \mbox{GeV}$ and $50\ \mbox{GeV}$, the $1S-2S$ energy
difference between $10\ \mbox{GeV}$ and $30\ \mbox{GeV}$. For the case of
gluinos being lighter than squarks the gluino decay rate is
significantly smaller than the level spacing and the boundstates are
thus well separated from the continuum. Also the wave function at the
origin, which enters bound state production and decay rates, has been
evaluated in NNLO. For $m_{\tilde{g}}>m_{\tilde{q}}$ the decay rate of
gluinonia into two gluon jets is evaluated in NLO and shown to dominate
all other modes. For top squarks with masses comparable to
$m_{\tilde{g}}$ the decay mode into $t\overline{t}$ may reach several
percent, the decay through virtual quark loops into two photons is too
small to be detectable. We evaluate the boundstate production cross
section again in NLO. Large positive corrections of around $50\%$ arise
from the NNLO terms in the wave function at the origin and around $70\%$
from the NLO terms in the cross section. The production rate depends
strongly on the gluino mass, with up to $10^6$ events for gluino masses
close to the lower limit around $300~\mbox{GeV}$ (for an integrated
luminosity of $100~\mbox{fb}^{-1}$) dropping quickly to $10$ events
only for $m_{\tilde{g}}=1.5~\mbox{TeV}$. Using standard assumptions for
the dijet-mass resolution and assuming that gluon and quark jets can be
separated, a signal-to-background ratio between $0.4$ and $0.7\%$ percent
might be conceivable. For favorable gluino- and squark-mass combinations
gluinonia might thus be detectable.


{\bf Acknowledgements}
M.S. thanks Alexander Penin for discussions about $f_2^{nC}$.
This work was supported by the BMBF through Grant No. 05H09VKE.


\medskip

{\it Note added}: After completion of this work we received a preprint by
K.~Hagiwara and H.~Yokoya where the production of
unstable gluinos in the threshold region is considered~\cite{Hagiwara:2009hq}.


\begin{appendix}

\section{\label{app:energy} Energy levels and wave functions}
For completeness we present the perturbative analytic results for the
energy levels $E_{n}$ of the $S$ waves and the corresponding values of the
squared wave function at the origin
$\left|\Psi_{n}\left(0\right)\right|^2$, truncating the perturbative
series at NNLO. We present the results following the notation of
\cite{Beneke:2005hg}. We define
\begin{eqnarray}
E_{n}&=&E_{n}^{(0)}\left[1+\frac{\alpha_{s}}{4\pi}e_{1}+\left(\frac{\alpha_{s}}{4\pi}\right)^2e_{2}\right]
\,,\nonumber\\
|\Psi_{n}(0)|^2&=&|\Psi_{n}^{(0)}(0)|^2\left[1+\frac{\alpha_{s}}{4\pi}f_{1}+\left(\frac{\alpha_{s}}{4\pi}\right)^2f_{2}\right]
\,,
\label{EundPsis}
\end{eqnarray}
and parameterize the corrections as
\begin{eqnarray}
e_{i}&=&e_{i}^{C}+e_{i}^{nC}\,,\nonumber\\
f_{i}&=&f_{i}^{C}+f_{i}^{nC}
\,,
\label{cundnc}
\end{eqnarray}
where $C$ stands for the corrections to the Coulombic part of the
potential and $nC$ for the remaining one. The LO values are given as
\begin{eqnarray}
E_{n}^{(0)}&=&-\frac{m_{\tilde{g}}C_{A}^2\alpha_{s}^2}{4 n^2}
\,,\nonumber \\
|\Psi_{n}^{(0)}(0)|^2&=&\frac{m_{\tilde{g}}^3C_{A}^3\alpha_{s}^3}{8\pi n^3}
\,.
\label{E0undPsi0}
\end{eqnarray}
The NLO and NNLO corrections are
\begin{eqnarray}
e_{1}^{C}&=&4\beta_{0}\ln\left(\frac{n\mu}{m_{\tilde{g}}C_{A}\alpha_{s}}\right)+2a_{1}+4S_{1}\beta_{0}
\,,\nonumber\\
e_{1}^{nC}&=&0
\,,\nonumber\\
e_{2}^{C}&=&12\beta_{0}^2\ln^2\left(\frac{n\mu}{m_{\tilde{g}}C_{A}\alpha_{s}}\right)+\ln\left(\frac{n\mu}{m_{\tilde{g}}C_{A}\alpha_{s}}\right)\left[-8\beta_{0}^2+4\beta_{1}+6\beta_{0}\left(2a_{1}+4S_{1}\beta_{0}\right)\right]
\nonumber\\
&&+a_{1}^2+2a_{2}+4S_{1}\beta_{1}+4a_{1}\beta_{0}\left(3S_{1}-1\right)
\nonumber\\
&&+\beta_{0}^2\left[S_{1}\left(12S_{1}-8-\frac{8}{n}\right)+16S_{2}-8nS_{3}+\frac{2\pi^2}{3}+8n\zeta(3)\right]
\,,\nonumber\\
e_{2}^{nC}&=&\frac{16\pi^2C_{A}^2}{n}\left(3-\frac{11}{16n}-\frac{2}{3}\vec{S}^2\right)
\,,\nonumber\\
f_{1}^{C}&=&6\beta_{0}\ln\left(\frac{n\mu}{m_{\tilde{g}}C_{A}\alpha_{s}}\right)+3a_{1}+2\beta_{0}\left(S_{1}+2nS_{2}-1-\frac{n\pi^2}{3}\right)
\,,\nonumber\\
f_{1}^{nC}&=&0
\,,\nonumber\\
f_{2}^{C}&=&24\beta_{0}^2\ln^2\left(\frac{n\mu}{m_{\tilde{g}}C_{A}\alpha_{s}}\right)
\nonumber\\
&&+\ln\left(\frac{n\mu}{m_{\tilde{g}}C_{A}\alpha_{s}}\right)\biggl\{-12\beta_{0}^2+6\beta_{1}+8\beta_{0}\left[3a_{1}+2\beta_{0}\left(S_{1}+2nS_{2}-1-\frac{n\pi^2}{3}\right)\right]\biggr\}
\nonumber\\
&&+3a_{1}^2+3a_{2}+2a_{1}\beta_{0}\left(4S_{1}+8nS_{2}-7-\frac{4n\pi^2}{3}\right)+2\beta_{1}\left(S_{1}+2nS_{2}-1-\frac{n\pi^2}{3}\right)
\nonumber\\
&&+\beta_{0}^2\biggl[S_{1}\left(8S_{1}+16nS_{2}-20-\frac{12}{n}-\frac{8n\pi^2}{3}\right)+S_{2}\left(4n^2S_{2}+8-8n-\frac{4n^2\pi^2}{3}\right) \biggr.
\nonumber\\
&&\hspace{0.3cm}\biggl.+28nS_{3}-20n^2S_{4}-24nS_{2,1}+16n^2S_{3,1}+4+\frac{(3+4n)\pi^2}{3}+\frac{n^2\pi^4}{9}+20n\zeta(3)\biggr]
\,,\nonumber\\
f_{2}^{nC}&=&16\pi^2C_{A}^2\biggl\{3\ln\left(\frac{n\mu}{m_{\tilde{g}}C_{A}\alpha_{s}}\right)-3S_{1}+\frac{6}{n}-\frac{15}{8n^2}+\frac{21}{4}\biggr.
\nonumber\\
&&\hspace{1.7cm}+\biggl.\vec{S}^2\left[-\frac{2}{3}\ln\left(\frac{n\mu}{m_{\tilde{g}}C_{A}\alpha_{s}}\right)+\frac{2}{3}S_{1}-\frac{4}{3n}-\frac{7}{9}\right]\biggr\}
\,.\label{eundf}
\end{eqnarray}
The constants $a_{1},a_{2},\beta_{0}$ and $\beta_{1}$ depend on the Casimir
operators $C_{A},C_{F}$ and $T_{F}$ and are defined in Eq.~(\ref{BetaundA}). The harmonic
sums $S_{i}(n)$, the nested harmonic sums $S_{i,j}(n)$ and the zeta-function
$\zeta(i)$ are given as
\begin{eqnarray}
  S_{i}\,\,=\,\,S_{i}(n)&=&\sum_{k=1}^{n}\frac{1}{k^i}
  \,,\nonumber\\
  S_{i,j}\,\,=\,\,S_{i,j}(n)&=&\sum_{k=1}^{n}\frac{1}{k^i}S_{j}(k)
  \,,\nonumber\\
  \hspace{0.45cm}\zeta(i)&=&\sum_{k=1}^{\infty}\frac{1}{k^{i}}
  \,.
  \label{S}
\end{eqnarray}
\section{\label{app:ttbargamgam} Amplitudes for the decays into top
  quarks and into photons}

\subsection{\label{app:ttbar} Decay into top quarks}

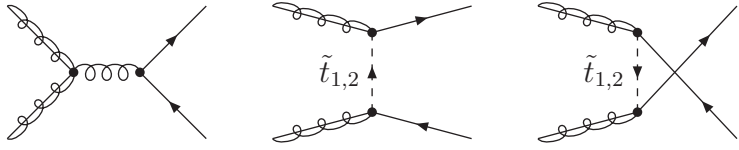
\begin{figure}[t]
\begin{center}
\begin{picture}(275,50)(0,0)
\SetColor{Black}
\Gluon(0,0)(25,25){-2.5}{3}
\Line(0,0)(25,25)
\Gluon(0,50)(25,25){2.5}{3}
\Line(0,50)(25,25)
\Vertex(25,25){1.8}
\Gluon(25,25)(50,25){3}{3}
\Vertex(50,25){1.8}
\ArrowLine(75,0)(50,25)
\ArrowLine(50,25)(75,50)
\Gluon(100,0)(137.5,10){-2.5}{3}
\Line(100,0)(137.5,10)
\Gluon(100,50)(137.5,40){2.5}{3}
\Line(100,50)(137.5,40)
\Vertex(137.5,40){1.8}
\DashArrowLine(137.5,10)(137.5,40){3}
\Vertex(137.5,10){1.8}
\ArrowLine(137.5,40)(175,50)
\ArrowLine(175,0)(137.5,10)
\Text(118,25)[l]{$\tilde{t}_{1,2}$}
\Gluon(200,0)(237.5,10){-2.5}{3}
\Line(200,0)(237.5,10)
\Gluon(200,50)(237.5,40){2.5}{3}
\Line(200,50)(237.5,40)
\Vertex(237.5,40){1.8}
\DashArrowLine(237.5,40)(237.5,10){3}
\Vertex(237.5,10){1.8}
\Line(237.5,10)(251.5625,25)
\Line(237.5,40)(251.5625,25)
\ArrowLine(251.5625,25)(275,50)
\ArrowLine(275,0)(251.5625,25)
\Text(218,25)[l]{$\tilde{t}_{1,2}$}
\end{picture}
\caption{Feynman diagrams for the decay of $(\tilde{g}\tilde{g})$ into
  $t\overline{t}$.}
\label{qqBAR}
\end{center}
\end{figure}

In Fig.~\ref{qqBAR} the corresponding diagrams for the decay of two
gluinos into top-antitop are shown. Diagram $(\alpha)$ drops out for
the case of interest because the two gluinos are projected here onto a
color-singlet state. The resulting amplitude for $(\beta)$ and
$(\gamma)$ reads
\begin{eqnarray}
\left.\mathcal{A}\right|_{\mbox{\tiny color
    singlet}}&=&\sum_{h=1}^{2}\frac{C_{F}\delta_{ij}}{\sqrt{N_{C}^2-1}}\frac{2ig_{s}^2}{m_{\tilde{g}}^2+m_{\tilde{t}_{h}}^2-m_{t}^2}\overline{u}_{s_{1}}(k_{1},m_{t})\left(P_{-}U_{h2}^{(\tilde{t})}-P_{+}U_{h1}^{(\tilde{t})}\right)
\nonumber\\
&&\times\left[\hspace{0.2cm}u_{\overline{s}}\left(\frac{P}{2},m_{\tilde{g}}\right)\overline{v}_{s}\left(\frac{P}{2},m_{\tilde{g}}\right)\right.
\nonumber\\
&&\left.\hspace{0.4cm}-\hspace{0.1cm}u_{s}\left(\frac{P}{2},m_{\tilde{g}}\right)\overline{v}_{\overline{s}}\left(\frac{P}{2},m_{\tilde{g}}\right)\right]\left(P_{-}U_{h1}^{(\tilde{t})}-P_{+}U_{h2}^{(\tilde{t})}\right)v_{s_{2}}(k_{2},m_{t})
\,,\label{ttBARamp}
\end{eqnarray}
where $s_{1},s_{2},s$ and $\overline{s}$ are the spin indices of the
top, the anti-top and of the gluinos, $P_{-}$ and $P_{+}$ are the
left- and right-handed projectors as usual and the
$U_{mn}^{(\tilde{t})}$ are the elements of the orthogonal squark-mixing matrix
\begin{eqnarray}
U^{(\tilde{t})}=\left(\begin{array}{cc}U_{11}^{(\tilde{t})}&U_{12}^{(\tilde{t})}\\U_{21}^{(\tilde{t})}&U_{22}^{(\tilde{t})}\end{array}\right)
\,.\label{mixing}
\end{eqnarray}
The relative minus sign between the diagrams $(\beta)$ and $(\gamma)$ of
Fig.~\ref{qqBAR} stems from the interchange of two fermionic operators.

The amplitude in Eq.~(\ref{ttBARamp}) has now, following the methods explained in
Refs. \cite{Guberina:1980dc,Kuhn:1979bb}, to be projected onto the
spin-$0$ state using the Clebsch-Gordon coefficients
$\left<0,0;s,\overline{s}\right>$, where the squared brackets of
Eq. (\ref{ttBARamp}) become due to the antisymmetry of the spin function
\begin{eqnarray}
&&\sum_{s,\overline{s}}\left[u_{\overline{s}}\left(\frac{P}{2},m_{\tilde{g}}\right)\overline{v}_{s}\left(\frac{P}{2},m_{\tilde{g}}\right)-u_{s}\left(\frac{P}{2},m_{\tilde{g}}\right)\overline{v}_{\overline{s}}\left(\frac{P}{2},m_{\tilde{g}}\right)\right]\left<0,0;s,\overline{s}\right>
\nonumber\\
&=&2\sum_{s,\overline{s}}\left[u_{\overline{s}}\left(\frac{P}{2},m_{\tilde{g}}\right)\overline{v}_{s}\left(\frac{P}{2},m_{\tilde{g}}\right)\right]\left<0,0;s,\overline{s}\right>
\,.\label{spinproj}
\end{eqnarray}
Now the calculation can be done as for the case of the decay into two
gluons and the rate reads
\begin{eqnarray}
\hspace{-0.5cm}
\Gamma\left(0^{-+}\rightarrow
  t\overline{t}\right)&=&\frac{C_{F}\alpha_{s}^2\left|R(0)\right|^2}{2m_{\tilde{g}}^2}\sqrt{1-\frac{m_{t}^2}{m_{\tilde{g}}^2}}\left(\sum_{h=1}^{2}\frac{m_{\tilde{g}}\left(m_{t}-2m_{\tilde{g}}U_{h1}^{(\tilde{t})}U_{h2}^{(\tilde{t})}\right)}{m_{\tilde{g}}^2+m_{\tilde{t}_{h}}^2-m_{t}^2}\right)^2
\,.\label{decayQQ}
\end{eqnarray}

\subsection{\label{app:gamgam} Decay into photons}

\begin{figure}[t]
\begin{center}
\SetScale{.9}
\begin{picture}(455,60)(-8,0)
\SetColor{Black}
\Gluon(0,0)(30,10){-2.5}{3}
\Line(0,0)(30,10)
\Gluon(0,55)(30,45){2.5}{3}
\Line(0,55)(30,45)
\Vertex(30,10){1.8}
\Vertex(30,45){1.8}
\DashArrowLine(30,10)(30,45){3}
\ArrowLine(65,10)(30,10)
\ArrowLine(30,45)(65,45)
\ArrowLine(65,45)(65,10)
\Vertex(65,10){1.8}
\Vertex(65,45){1.8}
\Photon(65,10)(95,0){-3}{3}
\Photon(65,45)(95,55){3}{3}
\Gluon(120,0)(150,10){-2.5}{3}
\Line(120,0)(150,10)
\Gluon(120,55)(150,45){2.5}{3}
\Line(120,55)(150,45)
\Vertex(150,10){1.8}
\Vertex(150,45){1.8}
\ArrowLine(150,45)(150,10)
\DashArrowLine(150,10)(185,10){3}
\DashArrowLine(185,45)(150,45){3}
\DashArrowLine(185,10)(185,45){3}
\Vertex(185,10){1.8}
\Vertex(185,45){1.8}
\Photon(185,10)(215,0){-3}{3}
\Photon(185,45)(215,55){3}{3}
\Gluon(240,-5)(305,10){-2.5}{6}
\Line(240,-5)(305,10)
\Gluon(240,55)(270,45){2.5}{3}
\Line(240,55)(270,45)
\Vertex(270,10){1.8}
\Vertex(270,45){1.8}
\DashArrowLine(270,10)(270,45){3}
\DashArrowLine(305,10)(270,10){3}
\ArrowLine(270,45)(305,45)
\ArrowLine(305,45)(305,10)
\Vertex(305,10){1.8}
\Vertex(305,45){1.8}
\Photon(270,10)(335,-5){-3}{7}
\Photon(305,45)(335,55){3}{3}
\Gluon(360,0)(390,10){-2.5}{3}
\Line(360,0)(390,10)
\Gluon(360,55)(390,45){2.5}{3}
\Line(360,55)(390,45)
\Vertex(390,10){1.8}
\Vertex(390,45){1.8}
\ArrowLine(390,10)(390,45)
\DashArrowLine(425,27.5)(390,10){3}
\DashArrowLine(390,45)(425,27.5){3}
\Vertex(425,27.5){1.8}
\Photon(425,27.5)(455,0){-3}{4}
\Photon(425,27.5)(455,55){3}{4}
\end{picture}
\caption{Representative diagrams for the decay into two photons at LO.}
\label{2gluto2gam}
\end{center}
\end{figure}
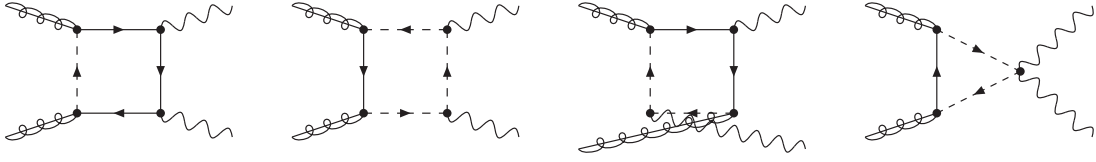

Treating all the virtual quarks in Fig.~\ref{2gluto2gam} as massless,
the onto spin and color-projected amplitude has the form
\begin{eqnarray}
&&\mathcal{A}((\tilde{g}\tilde{g})_{1_{S}}\rightarrow\gamma\gamma)
\nonumber\\
&=&-iT_{F}\sqrt{N_{C}^2-1}\frac{e^2g^2}{\sqrt{2}\pi^2}\varepsilon_{\alpha\beta\gamma\delta}k_{1}^{\alpha}k_{2}^{\beta}\epsilon_{1}^{\gamma}\epsilon_{2}^{\delta}\sum_{f,h}Q_{f}^2
\nonumber\\
&\times&\Biggl\{\frac{C_{0}(0,-m_{\tilde{g}}^2,m_{\tilde{g}}^2,m_{q_{f}}^2,m_{q_{f}}^2,m_{\tilde{q}_{f,h}}^2)-C_{0}(4m_{\tilde{g}}^2,0,0,m_{q_{f}}^2,m_{q_{f}}^2,m_{q_{f}}^2)}{m_{\tilde{g}}^2+m_{\tilde{q}_{f,h}}^2-m_{q_{f}}^2}\Biggr.\nonumber\\
&&\hspace{0.5cm}\times\left(m_{q_{f}}-2U^{(\tilde{f})}_{h1}U^{(\tilde{f})}_{h2}m_{\tilde{g}}\right)m_{q_{f}}
\nonumber\\
&&\hspace{0.4cm}\Biggl.+\frac{m_{\tilde{q}_{f,h}}^2C_{0}(0,-m_{\tilde{g}}^2,m_{\tilde{g}}^2,m_{\tilde{q}_{f,h}}^2,m_{\tilde{q}_{f,h}}^2,m_{q_{f}}^2)-m_{q_{f}}^2C_{0}(0,-m_{\tilde{g}}^2,m_{\tilde{g}}^2,m_{q_{f}}^2,m_{q_{f}}^2,m_{\tilde{q}_{f,h}}^2)}{2(m_{\tilde{q}_{f,h}}^2-m_{q_{f}}^2)}\Biggr\}\,.
\nonumber\\
\label{ampgamgam}
\end{eqnarray}
The sum in the first equation of (\ref{ampgamgam}) goes over the
flavors ($f$) and the indices of the squarks ($h$). The $k_{i}$ are the
momenta of the outgoing photons, the $\epsilon_{i}$ their
polarization vectors and the $Q_{f}$ the charges of the corresponding
quarks and squarks. The definition of the $C_{0}$ -functions is the same
as in Ref.~\cite{'tHooft:1978xw}.

\end{appendix}


\end{document}